\documentclass[11pt,english,onecolumn]{IEEEtran}
\usepackage[T1]{fontenc}
\usepackage[latin9]{inputenc}
\usepackage{geometry}
\usepackage{units}
\usepackage{amsmath}
\usepackage{amsthm}
\usepackage{amssymb}
\usepackage{graphicx}
\usepackage{setspace}
\makeatletter
\theoremstyle{plain}
\newtheorem{thm}{\protect\theoremname}
\theoremstyle{remark}
\newtheorem{rem}[thm]{\protect\remarkname}
\theoremstyle{plain}
\newtheorem{cor}[thm]{\protect\corollaryname}
\theoremstyle{definition}
\newtheorem{example}[thm]{\protect\examplename}
\theoremstyle{plain}
\newtheorem{prop}[thm]{\protect\propositionname}
\theoremstyle{definition}
\newtheorem{defn}[thm]{\protect\definitionname}
\theoremstyle{plain}
\newtheorem{lem}[thm]{\protect\lemmaname}
\theoremstyle{plain}
\newtheorem{fact}[thm]{\protect\factname}

\usepackage{hyperref}
\usepackage{enumitem}
\usepackage{breqn}
\usepackage{bbm} 
\usepackage{cite}

\DeclareMathOperator{\Cat}{Categorical} 
\DeclareMathOperator{\Mul}{Multinomial} 
\DeclareMathOperator{\Poi}{Poisson} 
\DeclareMathOperator{\Ber}{Bernoulli} 
\DeclareMathOperator{\Gam}{Gamma} 
 
\DeclareMathOperator{\Uni}{Uniform} 
 
\DeclareMathOperator{\Geo}{Geo} 
\DeclareMathOperator{\supp}{supp} 
\DeclareMathOperator{\Ent}{Ent} 
\DeclareMathOperator{\mmpe}{mmpe} 
\DeclareMathOperator{\erf}{erf} 
\DeclareMathOperator{\Q}{Q} 
\DeclareMathOperator{\W}{W}

\global\long\def\P{\mathbb{P}}
\global\long\def\E{\mathbb{E}}
\global\long\def\V{\mathrm{Var}}

\global\long\def\I{\mathbbm{1}}
\global\long\def\d{\mathrm{d}}

\global\long\def\Dkl{\mathrm{D_{KL}}}
\global\long\def\hbin{\mathrm{h_{bin}}}

\global\long\def\eqd{\stackrel{d}{=}}

\global\long\def\trre[#1,#2]{\overset{{\scriptstyle (#2)}}{#1}} 
\renewcommand\[{\begin{equation}}
\renewcommand\]{\end{equation}}

\allowdisplaybreaks

\author{Yuval Gerzon$^{1}$, Ilan Shomorony$^{2}$ and Nir Weinberger$^{1}$}

\makeatother

\usepackage{babel}
\providecommand{\corollaryname}{Corollary}
\providecommand{\definitionname}{Definition}
\providecommand{\examplename}{Example}
\providecommand{\factname}{Fact}
\providecommand{\lemmaname}{Lemma}
\providecommand{\propositionname}{Proposition}
\providecommand{\remarkname}{Remark}
\providecommand{\theoremname}{Theorem}

\begin{document}
\title{Capacity of Frequency-based Channels: Encoding Information in Molecular
Concentrations\thanks{$^{1}$ Yuval Gerzon and Nir Weinberger are with the Viterbi Faculty
of Electrical and Computer Engineering, Technion-Israel Institute
of Technology, Haifa 3200004, Israel (Emails: gerzon.yuval@campus.technion.ac.il,
nirwein@technion.ac.il.). $^{2}$ Ilan Shomorony is with the Department
of Electrical and Computer Engineering, University of Illinois at
Urbana-Champaign, Urbana, IL 61801 USA (Email: ilans@illinois.edu).
This paper was accepted in part to the 2024 IEEE International Symposium
on Information Theory. The research of N. W. was supported by the
Israel Science Foundation (ISF), grant no. 1782/22. The work of I.
S. was supported in part by the National Science Foundation under
CCF grants 2007597 and 2046991.}}
\maketitle
\begin{abstract}
We consider a molecular channel, in which messages are encoded to
the frequency of objects (or concentration of molecules) in a pool,
and whose output during reading time is a noisy version of the input
frequencies, as obtained by sampling with replacement from the pool.
We tightly characterize the capacity of this channel using upper and
lower bounds, when the number of objects in the pool of objects is
constrained. We apply this result to the DNA storage channel in the
short-molecule regime, and show that even though the capacity of this
channel is technically zero, it can still achieve a large information
density. 
\end{abstract}


\section{Introduction}

In molecular communication \cite{gohari2016information}, information
is encoded into the presence of objects from various possible types
in some restricted physical domain. As a prominent example, in DNA
storage systems \cite{church2012next,goldman2013towards,grass2015robust,yazdi2015rewritable,kiah2016codes,erlich2017dna,sala2017exact,organick2018random,lenz2019anchor,sima2021coding,tang2021error},
information is encoded to $K$ molecules, each is a strand of length
$L$ composed from the four possible nucleotides, denoted by ${\cal A}:=\{\text{A},\text{C},\text{G},\text{T}\}$.
The $K$ molecules are stored in a pool, and the distinctive aspect
of this system is that the order of the $K$ molecules in the pool
cannot be preserved. So, while the total of $KL$ symbols can be considered
as the blocklength of the codeword, unlike standard channel coding,
the codeword is, in fact, partitioned to $K$ out-of-order segments
of length $L$ each. Given the pool, the message is decoded by randomly
sampling molecules from the pool, sequencing each of them to obtain
a noisy read of the sequence of nucleotides in the strand, and using
the out-of-order output strands to decode the message. 

A simple method to resolve the lack of order of the encoded molecules
is to devote the first $\log_{|{\cal A}|}K$ symbols of the molecule
to encode its index. This requires that the molecule length will be
large enough to accommodate both the index and the message encoding.
Indeed, in \cite{shomorony2021dna} it was established that the regime
of interest is $L=\beta\log K$ for some $\beta>0$, and it was shown
that indexing achieves the capacity of the DNA storage channel when
the sequencing is noiseless. Conversely, if the molecule length $L$
is short in comparison to their number $K$, and so there is not enough
symbols to encode the index in the molecule, concretely, if $\beta\leq\frac{1}{\log|{\cal A}|}$,
then the capacity is \emph{zero}. Consequently, in this regime, the
log-cardinality of the optimal codebook (which is the total number
of stored bits) scales at most \emph{sub-linearly} with the total
number of nucleotides $KL$. 

Nonetheless, the DNA storage medium has an extreme information density,
and a huge number of nucleotides can be stored in tiny pools. As noted
in \cite[Sec. 7.3]{shomorony2022information}, the amount of stored
information can be large even for channels with asymptotically vanishing
maximal rate. This observation serves as a strong motivation to study
the log-cardinality of the optimal codebook in the short-molecule
regime, that is, for $\beta\leq\frac{1}{\log|{\cal A}|}$. In this
regime, the number of molecules in a codeword $K$ is larger than
the number of possible strands of length $L$ from the alphabet ${\cal A}$,
to wit, $K\geq|{\cal A}|^{L}$. Accordingly, each codeword must contain
multiple copies of the same strand (at least for one strand). Hence,
in the short-molecule regime, the message is actually encoded by the
number of times each of the $|{\cal A}|^{L}$ possible strands of
length $L$ appears in the pool. The encoded codeword can thus be
represented by the frequency vector that measures the frequency of
each type of strands in the pool. We refer to this type of channel
as \emph{frequency-based channel}. During reading, molecules are sampled
from the pool (with replacement) and so the output is a also a frequency
vector. This vector is a noisy version of the input vector for two
reasons: First, the frequency of the strands in the codeword is not
preserved by the sampling. Second, the synthesis and the sequencing
processes are possibly noisy \cite{sabary2021solqc}. 

In this paper, we formulate a general frequency-based channel. We
focus on the effect of random sampling, which we model as a multinomial
distribution, and thus assume noiseless sequencing. In this channel
model, the blocklength $n$ models the number of different types of
objects. Each codeword has a total count of $ng_{n}$ objects from
the different types, and which are sampled $nr_{n}$ times in total.
For example, in the DNA storage channel $n=|{\cal A}|^{L}$, as this
is the number of different molecules of length $L$ from an alphabet
${\cal A}$, and $ng_{n}=K$, as this is the total number of molecules.
In \cite[Sec. 7.3]{shomorony2022information} a slightly different
Poisson sampling channel was considered, which assumes $g_{n}=r_{n}$,
and a conjecture was made on the scaling of the log-cardinality of
the optimal codebook \cite[Conjecture 4]{shomorony2022information}.
Concretely, based on the capacity of the average-power-constrained
Poisson channel \cite{lapidoth2008capacity} it was conjectured that
the capacity scales as $\frac{1}{2}\log r_{n}+o_{n}(1)$. 

\paragraph*{Main contribution}

In this work, we address that conjecture. Though our multinomial sampling
is slightly different, its analysis is, in fact, based on a reduction
to a Poisson channel, and so in this sense the multinomial model subsumes
the Poisson model. That being said, the reduction itself complicates
the analysis of the resulting Poisson channel, and the latter is non-standard
from two aspects. First, the total number of counts in the codeword
must be common to all codewords in the codebook in order for an input
count to accurately model frequencies. Second, since the input symbols
also measure the count of a possible objects in the codeword, they
must be \emph{integer-valued}, and so the input distribution must
be supported on the integers. Our main result (Theorem \ref{thm: noiseless kernel bounds})
is an approximate solution of the conjecture of \cite[Conjecture 4]{shomorony2022information}:
Our converse bound shows that the capacity is less then $\frac{1}{2}\log[r_{n}\wedge(eg_{n})]+o_{n}(1)$.
That is, increasing $r_{n}$ beyond $g_{n}$ may increase capacity,
but asymptotically only up to $\frac{1}{2}[\text{nats}]$. Our achievable
bound requires the condition $n=\omega(g_{n})$, and when the ratio
$r_{n}/g_{n}$ is optimized, it is given by $\frac{1}{2}\log(g_{n})-1.295+o_{n}(1)[\text{nats}]$.
Interestingly, the optimum of the lower bound occurs at $r_{n}\approx0.4g_{n}$,
i.e., when sampling \emph{less} objects than there are in the codeword. 

The implication of this result to the DNA storage channel is valid
when the molecules are not very short, and specifically, in the regime
$\beta\in(\frac{1}{2\log|{\cal A}|},\frac{1}{\log|{\cal A}|})$. The
result shows that the log-cardinality of the optimal codebook increases
as 
\[
\frac{1-\beta\log|{\cal A}|}{2\beta}\cdot K^{\beta\log|{\cal A}|}\log K,
\]
up to terms negligible with $K$. A simple numerical example then
shows that the resulting information density (in nats per gram) could
still be huge, which remarkably reinforces the importance of this,
strictly speaking, zero capacity, channel. 

\paragraph*{Related work}

The analysis of the DNA-storage channel is an active research area,
both from an information-theoretic point of view \cite{shomorony2021dna,lenz2019upper,lenz2020achieving,weinberger2022Error,weinberger2022dna,shomorony2022information,vippathalla2023secure}
and from coding-theoretic point of view \cite{lenz2019anchor,lenz2019coding,lenz2020achievable,sima2021coding}.
Our channel model is closely-related to the \emph{permutation channel}.
Using our formulation, the input to the permutation channel is also
a frequency vector, however, it is assumed that each object is sampled
exactly once, and the output vector is noisy due to noisy sequencing
(in our terms). This channel was considered in \cite{kovavcevic2017coding,kovavcevic2018codes}
with codes termed \emph{multiset codes}. Constructions of such codes
were proposed, and combinatorial bounds on the size of optimal codes
for a given detection or correction capability were derived. An information-theoretic
version of the channel was introduced in \cite{makur2020coding},
with sharp converse bounds obtained in \cite{tang2023capacity}. However,
compared to our results, and in our terminology, the blocklength in
\cite{makur2020coding,tang2023capacity} is considered fixed, whereas
in our model it increases without bound (with a certain scaling).
A multi-user model of this channel was recently explored in \cite{lu2024permutation}.
Another related model for DNA storage is based on composite DNA letters
\cite{choi2019high,anavy2019data}, in which many copies of a single
molecule are generated, and each letter in the molecule is a \emph{composite
letter}, i.e., it is randomly chosen from a subset of $\{\text{A},\text{C},\text{G},\text{T}\}$,
chosen according to the encoded information. In this channel model
too, information is stored in the frequency of each DNA letter at
the output, though the randomness of each letter is created during
synthesis. This leads to a somewhat different mathematical model,
of a multinomial channel, and its capacity is discussed, e.g., in
\cite{kobovich2023m}.

We rely on the analysis of the Poisson channel under an average-power
constraint. The capacity of this channel was extensively explored
(e.g., \cite{frey1991information,brady1990asymptotic,shamai1993bounds,lapidoth1998poisson,martinez2007spectral,lapidoth2008capacity,atar2012mutual,cao2013capacity,cao2013capacityII,cheraghchi2019improved,dytso2020vector,dytso2021properties}),
but for the frequency-based channel we mainly rely on the asymptotic
expression of \cite{lapidoth2008capacity}. On its own, the entropy
of a Poisson random variable (RV) was also extensively explored \cite{harremoes2001binomial,adell2010sharp,sason2013entropy},
along with ample study of related properties, e.g., \cite{harremoes2001binomial,kontoyiannis2005entropy,johnson2007log,yu2009monotonic,wang2014bregman,dytso2020estimation,dytso2020vector}. 

\paragraph*{Paper outline}

The paper is organized as follows. In Sec. \ref{sec:Problem-Formulation}
we shortly describe the DNA storage channel, and then formulate the
more general frequency-based channel. In Sec. \ref{sec:Main-Result}
we state our main result and outline its proof. In Sec. \ref{sec:Proofs}
we provide detailed proofs, and in Sec. \ref{sec:Conclusion-and-Future}
we conclude the paper with a summary and future research directions.

\section{Problem Formulation \label{sec:Problem-Formulation}}

\paragraph*{Notation conventions}

For an integer $n\in\mathbb{N}_{+}$, let $[n]:=\{1,2,\ldots n\}$.
For $a,b\in\mathbb{R}$, let $a\vee b:=\max\{a,b\}$ and $a\wedge b:=\min\{a,b\}$.
Logarithms and exponents are taken to the natural base. Standard notation
for information-theoretic quantities is used \cite{cover2012elements},
e.g., the entropy $H(P_{X})$ or $H(X)$ for a discrete RV $X$ with
probability mass function (PMF) $P_{X}$, the mutual information $I(X;Y)$
between two RVs $X$ and $Y$, and $\Dkl(P\mid\mid Q)$ for the KL
divergence between the probability measures $P$ and $Q$. The binary
entropy function is denoted by $\hbin(\cdot)$. 

Although our results are valid under general molecular storage settings,
the DNA storage channel is our main motivation, and so we next briefly
review its model, and explain how it translates in the short-molecule
regime to a frequency-based channel \cite[Sec. 7.3]{shomorony2022information}. 

\subsection{The DNA Storage Channel Model \label{subsec:DNA storage model}}

The DNA storage model, also called the \emph{multi-draw noisy shuffling
channel} \cite{shomorony2022information}, is as follows. Let an alphabet
${\cal A}$ be given, e.g., ${\cal A}=\{\text{A},\text{C},\text{G},\text{T}\}$
in the case of an actual DNA-based storage system. A codeword is $a^{LK}=(a_{1}^{L},\ldots a_{K}^{L})$,
where $a_{k}^{L}\in{\cal A}^{L}$ for all $k\in[K]$ is called a \emph{molecule}
or a \emph{strand}. Thus, there are $K$ molecules in a codeword,
each of which is a length-$L$ vector from the alphabet ${\cal A}$.
A codebook is a set of different codewords, ${\cal C}=\{a^{LK}(j)\}_{j\in[M]}$.
The codeword is read in a noisy way comprised of two stages. In the
first stage, $N$ molecules are uniformly sampled from the $K$ molecules
of $a^{LK}$, independently, with replacement. Letting $\{U_{i}\}_{i\in[N]}$
be independent and identically distributed (IID) such that $U_{i}\sim\Uni[K]$,
the output of this stage is $\{a_{U_{1}}^{L},\ldots,a_{U_{N}}^{L}\}$.
In the second stage, each of the sampled molecules $a_{U_{i}}^{L}$
is sequenced, and the output molecule $b_{i}^{*}\in{\cal B}^{*}$
is obtained, where ${\cal B}$ is an output alphabet, and $*$ indicates
varying length ${\cal B}^{*}=\bigcup_{\ell=1}^{\infty}{\cal B}^{\ell}$.
Thus, the length of $b_{i}^{*}$ may be different from $L$, and vary
from one molecule to the other. The possibly noisy sequencing is modeled
as a noisy channel $V_{L}$ from ${\cal A}^{L}$ to ${\cal B}^{*}$.
For example, this channel may include \emph{substitutions} of a letter
from ${\cal A}$ with a different letter, deletions, and insertions
\cite{heckel2019characterization}. The channel output is then $(b_{1}^{*},\ldots,b_{N}^{*})$.
\begin{rem}
It may seem more natural to model the channel output as obtained via
sampling \emph{without} replacement, since each sampled molecule is
removed from the DNA pool in order to be sequenced. The reason why
we model the channel as performing sampling \emph{with} replacement
is because in practical DNA storage systems, \emph{polymerase chain
reaction} (PCR) amplification is used to replicate each molecule in
the pool a large (but roughly fixed) number of times. Hence, the relative
frequency of each DNA molecule in the pool remains roughly fixed,
but sampling from this amplified DNA pool is essentially sampling
with replacement from the original pool.
\end{rem}
We let the length of each molecule scale as $L=L_{K}$, the number
of sampled molecules to scale as $N=N_{K},$ and denote the maximal
cardinality of a codebook with codewords of $K$ molecules and maximal
error probability $\epsilon_{K}$ as $M_{\text{DNA}}^{*}(L_{K},V_{L_{K}},N_{K},\epsilon_{K})$.
As a codeword is composed from a total $KL$ symbols from ${\cal A}$,
the rate of a codebook of cardinality $M$ is defined as $R:=\frac{1}{KL}\log M$.
The capacity is the maximal rate with vanishing error probability,
that is, $\frac{1}{KL}\log M_{\text{DNA}}^{*}(L_{K},V_{L_{K}},N_{K},\epsilon_{K})$
with $\epsilon_{K}\to0$ as $K\to\infty$. 

\paragraph*{The short-molecule regime}

The capacity of the DNA storage channel for discrete memoryless sequencing
channels was analyzed in \cite{shomorony2021dna,lenz2019upper,lenz2020achieving,weinberger2022dna}.
Specifically, it was shown in \cite{shomorony2021dna} that even for
\emph{noiseless} sequencing channels, the capacity of DNA storage
is strictly positive only if the molecule length scales as $L_{K}=\beta\log K$
and $\beta>\frac{1}{\log|{\cal A}|}$. Intuitively, the lack of order
in the codeword can be resolved by using $\log_{|{\cal A}|}K=\frac{\log K}{\log|{\cal A}|}$
symbols from each molecule to encode its index in the codeword, and
using the rest of the symbols to encode the message. However, if $\beta<\frac{1}{\log|{\cal A}|}$
then the length of a molecule does not allow for encoding the index,
let alone for encoding the message. It can be shown that indexing
is optimal for noiseless sequencing channels (though not necessarily
for noisy channels), and that no positive rate can be achieved when
$\beta\leq\frac{1}{\log|{\cal A}|}$. We thus refer to this regime
as the \emph{short-molecule regime.} From a different point of view,
we may note that the number of distinct molecules of length $L$ is
$|{\cal A}|^{L}$. As $K\geq|{\cal A}|^{L}$ in the short-molecule
regime, each codeword must contain more than a single copy of the
same molecule. Since the molecules of the codeword lack any order,
the message is actually encoded in the number of copies of each of
the $n=|{\cal A}|^{L}$ possible molecules in ${\cal A}^{L}$, or
in their \emph{frequencies}. Let us order the ${\cal A}^{L}$ strings
representing the possible molecules in some arbitrary order $[n]$.
Then, the input codeword can be equivalently represented by the vector
$x^{n}:=(x_{1},\ldots,x_{n})\in\mathbb{N}^{n}$ where $x_{i}$ is
the count of the $i$th string in ${\cal A}^{L}$. Thus, it holds
that $\sum_{i=1}^{n}x_{i}=K$. Similarly, let us denote the number
of counts of each of the strings in ${\cal A}^{L}$ at the output
codeword by $y^{n}:=(y_{1},\ldots,y_{n})\in\mathbb{N}^{n}$. Due to
the randomness in the sampling stage, $y^{n}$ is a noisy version
of $x^{n}$ even for noiseless sequencing channels. We refer to this
equivalent channel model as a frequency-based channel, and formally
define it in the next subsection, in greater generality. In the rest
of the paper we will analyze the capacity of that channel model, which,
in turn, leads to bounds on $M_{\text{DNA}}^{*}(L_{K},V_{L_{K}},N_{K},\epsilon_{K})$
for DNA storage channels. The capacity is zero in the short-molecule
regime, and so $\log M_{\text{DNA}}^{*}(L_{K},V_{L_{K}},N_{K},\epsilon_{K})$
scales \emph{sub-linearly} with $KL$. However, it is still a monotonic
non-decreasing function of $K$, for which our goal it to characterize
the optimal asymptotic scaling.

\subsection{The Frequency-based Channel Model \label{subsec:The-Concentration-Input-Channel}}

\begin{figure}
\begin{centering}
\includegraphics[scale=0.75]{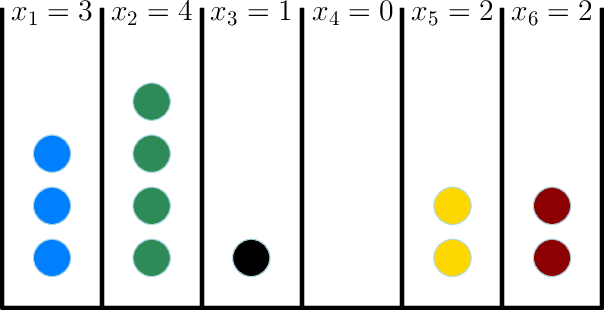}\\
~\\
\par\end{centering}
\begin{centering}
\includegraphics[scale=0.75]{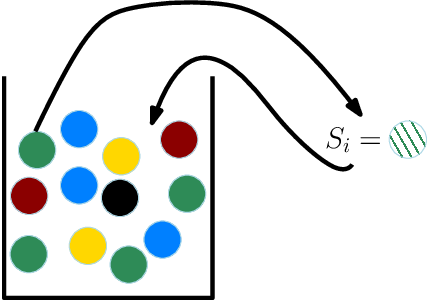}\\
~\\
\par\end{centering}
\centering{}\includegraphics[scale=0.75]{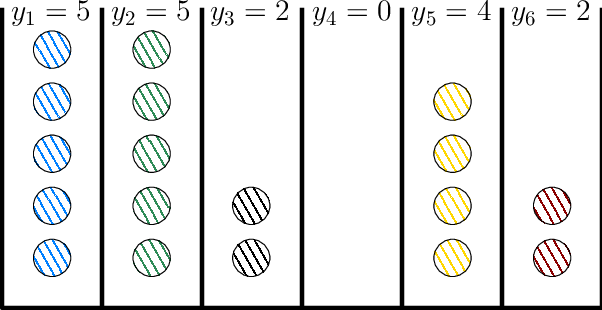}\caption{An illustration of the channel model with $n=6$, $g_{n}=2$ and $r_{n}=3$.
Top: The message is encoded to the codeword $x^{6}=(3,4,1,0,2,2)$.
Middle: The $ng_{n}=12$ objects are stored in a pool, and then sampled
with replacement $nr_{n}=18$ times. At each sample the object type
is recorded as $S_{i}$. Bottom: The output vector is the histogram
of $S^{nr_{n}}$, given by $y^{n}=(5,5,2,0,4,2)$. \label{fig: channel demonstration}}
\end{figure}

Consider a set of $n$ distinguishable types of objects (e.g., molecules).
An input message is encoded as a pool of unordered objects from the
various types. Thus, the channel input is represented by the count
vector $x^{n}:=(x_{1},\ldots,x_{n})\in\mathbb{N}^{n}$, where $x_{i}$
is the number of objects of the $i$th type in the pool of objects.
It is assumed that $\sum_{i=1}^{n}x_{i}$ is constant for all possible
messages. Thus, $\hat{x}_{i}:=x_{i}/(\sum_{i=1}^{n}x_{i})$ is a \emph{frequency-vector}
(or \emph{concentration}) of the $i$th type in the codeword pool.
It is further assumed that the total number of objects is restricted
as $\sum_{i=1}^{n}x_{i}\leq ng_{n}$, for some given $g_{n}$. To
read the message, $nr_{n}$ samples are taken, where for each $i\in[nr_{n}]$,
an object is randomly chosen uniformly at random from the set of $\sum_{i=1}^{n}x_{i}$
objects in the pool, with replacement. Then, the type of the object
is read, by a possibly noisy mechanism. Let $W_{n}\in\mathbb{R}^{n\times n}$
be a Markov kernel, so that $W_{n}(j,i)$ represents the probability
that an object of type $i$ is determined to be of type $j$ (hence
$W_{n}(j,i)\geq0$ and $\sum_{j=1}^{n}W_{n}(j,i)=1$). Thus, the $i$th
object is $S_{i}\sim\Cat(\hat{x}^{n}W_{n})$, where $\hat{x}^{n}W_{n}$
represents the standard multiplication of row vector by a matrix,
and $S^{nr_{n}}:=(S_{1},\cdots,S_{nr_{n}})$ is a vector of IID RVs.
Conditioned on input $x^{n}$, the output is equivalently a noisy
count vector $Y^{n}:=(Y_{1},\ldots,Y_{n})\in\mathbb{N}^{n}$ where
$Y^{n}\sim\Mul(nr_{n},\hat{x}^{n}W_{n})$. A noiseless setting is
illustrated in Fig. \ref{fig: channel demonstration}.

A code is a set of $M$ input count vectors ${\cal C}_{M}:=\{x^{n}(1),\ldots,x^{n}(M)\}$
for which $\sum_{i=1}^{n}x_{i}(m)$ is constant for all $m\in[M]$.
The size of the largest code for $n$ object types, normalized total
count of input objects $g_{n}$, normalized number of sampled objects
$r_{n}$, a reading kernel $W_{n}$, under a given error probability
$\epsilon_{n}\in(0,1)$ is denoted by $M^{*}(n\mid\epsilon_{n},g_{n},r_{n},W_{n})$.
Our goal is to accurately determine the growth rate of $M^{*}(n\mid\epsilon_{n},g_{n},r_{n},W_{n})$,
or the rate of the codebook, given by $\frac{1}{n}\log M^{*}(\cdot)$.
We assume that both $g_{n},r_{n}$ are monotonic non-decreasing functions
of $n$, and aim to accurately characterize the dependency of $M^{*}(n\mid\epsilon_{n},g_{n},r_{n},W_{n})$
on these sequences. For reasons that will be clear in what follows,
we focus on the regime $r_{n}=\Theta(g_{n})$. In this paper, we focus
on the randomness stemming from the sampling channel, and thus assume
that $W_{n}=I_{n}$, the noiseless kernel for all $n$ (we discuss
possible extensions to noisy channels in Sec. \ref{sec:Conclusion-and-Future}).
For the error probability, we just assume that $\epsilon_{n}\to0$
as $n\to\infty$, though in possibly arbitrarily slow rate. Our main
theorem provides upper and lower bounds on the rate $\frac{1}{n}\log M^{*}(\cdot)$
in this regime. 

In \cite[Sec. 7.3]{shomorony2022information}, a closely-related channel
model was considered, in which $r_{n}=g_{n}$, the sequencing is noiseless
$W_{n}=I_{n}$, and the output is $Z^{n}=(Z_{1},\ldots,Z_{n})$ is
such that conditioned on $X^{n}=x^{n}$, it holds that $Z_{i}\sim\Poi(x_{i})$,
and the $Z_{i}$'s are independent. Based on the known asymptotic
capacity expression of the Poisson channel under an average-power
constraint \cite[Thm. 7]{lapidoth2008capacity}, it was conjectured
that the capacity of the frequency-based channel whose output is $Z^{n}$
is given by $\frac{1}{2}\log r_{n}+o_{n}(1)$. However, the capacity
of the Poisson channel is asymptotically achieved by a gamma distributed
input $X_{n}\sim\Gam(\frac{1}{2},2g_{n})$ \cite{martinez2007spectral,lapidoth2008capacity},
which is a continuous distribution, whereas the frequency-based channel
only allows for integer inputs. Our bounds will be proved by modifying
the multinomial output to a Poisson output. However, the analysis
of the resulting Poisson channel, along with integer-input constraints
will require additional technical steps. 

\section{Main Result \label{sec:Main-Result}}

Assume that reading operation is noiseless, and so $W=I_{n}$. The
channel is given by 
\[
Y^{n}\sim\Mul\left(nr_{n},\frac{1}{\sum_{i=1}^{n}x_{i}}x^{n}\right).
\]
For $\mu\in\mathbb{R}_{+}$, let 
\begin{equation}
\Psi(\mu):=(\mu+1)\cdot\hbin\left(\frac{1}{\mu+1}\right),\label{eq: maximum entropy single letter}
\end{equation}
which, as is well known, is the maximum entropy for non-negative integer-supported
RVs with mean bounded by $\mu$ (see Lemma \ref{lem:Geometric max entropy under mean constraint}).
\begin{thm}
\label{thm: noiseless kernel bounds}Assume $W_{n}=I_{n}$, that $g_{n}\to\infty$,
and that $\underline{c}g_{n}\leq r_{n}\leq eg_{n}$ for some $\underline{c}\in(0,e)$.
\begin{itemize}
\item A (weak) converse bound: For any $\epsilon_{n}\to0$
\[
\frac{1}{n}\log M^{*}(n\mid\epsilon_{n},g_{n},r_{n},W_{n})\leq\frac{1}{2}\log\left[r_{n}\wedge(eg_{n})\right]+o_{n}(1).
\]
\item An achievability bound: Further assume that $n=\Omega(g_{n}^{1+\zeta})$
for some $\zeta>0$. Then, 
\[
\frac{1}{n}\log M^{*}(n\mid\epsilon_{n},g_{n},r_{n},W_{n})\geq\frac{1}{2}\log(r_{n})-\Psi\left(\frac{r_{n}}{g_{n}}\right)+o_{n}(1).
\]
\end{itemize}
\end{thm}

\paragraph*{Comparison to the standard Poisson channel}

For a Poisson channel with an average-power input constraint $\E[X]\leq g_{n}$,\footnote{For the Poisson channel, the average input power is modeled as $\E[X]$,
as opposed to the more common Gaussian channel, in which the average
input power is $\E[X^{2}]$.} and gain $\frac{r_{n}}{g_{n}}$, that is, $Z\mid X=x\sim\Poi(\frac{r_{n}}{g_{n}}x)$,
the capacity is asymptotically given by $\frac{1}{2}\log r_{n}+o_{n}(1)$,
\cite[Thm. 7]{lapidoth2008capacity}.\footnote{In \cite[Thm. 7]{lapidoth2008capacity}, the Poisson channel is assumed
to have unity gain, i.e., $\frac{r_{n}}{g_{n}}=1$. The capacity expression
can be easily generalized to non-unity gain by scaling of input codewords. } Nonetheless, this rate is achieved with input distribution $X\sim\Gam(\frac{1}{2},2g_{n})$,
which is a continuous distribution, and is unsuitable for the frequency-based
channel, which accepts non-negative \emph{integer} inputs. This restriction
on the input affects both the converse and the achievability bound.
For the converse part, it leads to an upper bound $\frac{1}{2}\log(eg_{n})$
on the maximal rate, which, as we discuss in what follows, is a result
of the log-cardinality of the set of possible inputs. Thus, unlike
the standard, continuous-input, Poisson channel, there is no motivation
to increase $r_{n}$ beyond $eg_{n}$, at least in terms of rate.
For the achievability bound, the restriction of the input to be integer
valued leads to the loss additive term $\Psi(\frac{r_{n}}{g_{n}})$.
This leads to a delicate issue: For the decoder, increasing $r_{n}$,
the number of samples from the pool of objects, only leads to higher
mutual information and rate, since due to the data-processing theorem,
the decoder can always ignore output observations. However, in our
model increasing $r_{n}$ also put a more restrictive constraint on
the input integer constraint, and this has the opposite effect on
the mutual information, as it does not allow to achieve the output
entropy obtained as in the standard case. Thus, it is not obvious
that increasing $r_{n}$ also increases the mutual information. To
further inspect this, let us write the lower bound, without the asymptotically
vanishing terms, as 
\[
\frac{1}{2}\log(g_{n})+\frac{1}{2}\log\left(\frac{r_{n}}{g_{n}}\right)-\Psi\left(\frac{r_{n}}{g_{n}}\right).
\]
We may then optimize it over $r_{n}\leq eg_{n}$. Interestingly, the
function $\mu\to\frac{1}{2}\log(\mu)-\Psi(\mu)$ has a unique global
maximum at $\mu\approx0.398$ and equals $-1.295[\text{nats}]$, which
is better than its value at $\mu=1$, given by $-1.386[\text{nats}]$.
Thus, to optimize the lower bound of Theorem \ref{thm: noiseless kernel bounds},
the optimal choice is $r_{n}\approx0.4g_{n}$, that is, \emph{surely}
not sampling some of the objects in the pool optimizes this bound.
The optimized lower bound is then (in nats)
\[
\frac{1}{n}\log M^{*}(n\mid\epsilon_{n},g_{n},r_{n},W_{n})\geq\frac{1}{2}\log(g_{n})-1.295+o_{n}(1).
\]
Naturally, an interesting open question is whether the rate can go
beyond $\frac{1}{2}\log(g_{n})$, and if it can match the upper bound,
and what is the optimal value of $r_{n}$. 

\paragraph*{Implication on the data stored in DNA storage systems}

\begin{figure}
\centering{}\includegraphics[scale=0.35]{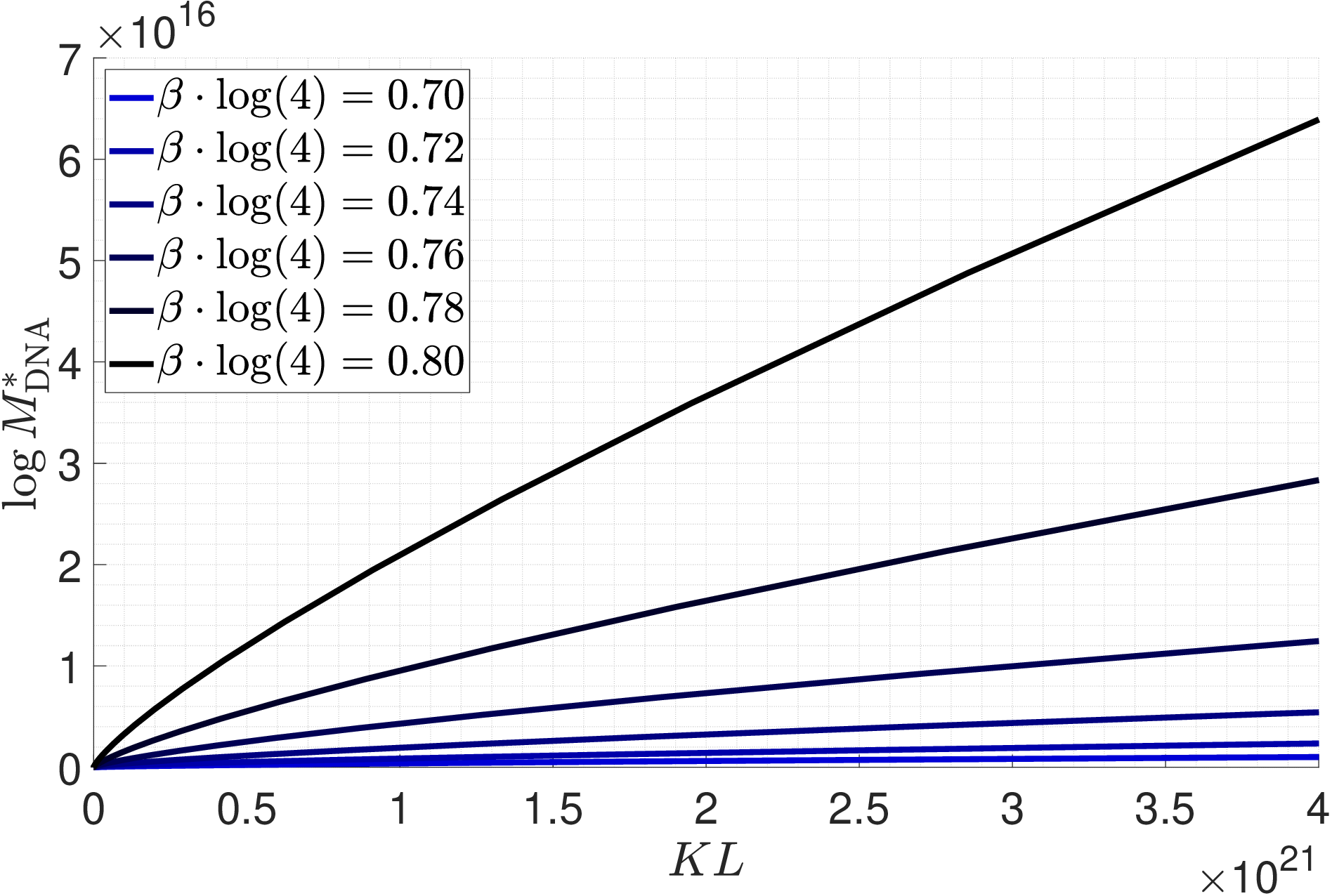}\caption{The lower bound on $\log M_{\text{DNA}}^{*}(L_{K},V_{L_{K}},N_{K},\epsilon_{K})$
of (\ref{eq: explicity lower bound DNA storage}) vs. $KL$. A darker
line corresponds to larger $\beta$. \label{fig: log cardinality lower bound versus KL}}
\end{figure}

As discussed, strictly speaking, the capacity of the DNA storage channel
in the regime of interest is zero. Following \cite[Sec. 7.1]{shomorony2022information},
let the pseudo-rate of a DNA storage be defined as 
\[
\tilde{R}_{\text{DNA}}:=\frac{\log M}{LK^{\beta\log|{\cal A}|}},
\]
and the pseudo-capacity be defined as the maximal achievable pseudo-rate
such that $\epsilon_{K}\to0$ as $K\to\infty$. We thus obtain the
following corollary:
\begin{cor}
\label{cor:DNA storage pseudo-rate}Assume that $\beta>\frac{1}{2\log|{\cal A}|}$.
Then, 
\[
\tilde{R}_{\text{\emph{DNA}}}=\frac{1-\beta\log|{\cal A}|}{2\beta}.
\]
\end{cor}
Specifically, this settles \cite[Conjecture 4]{shomorony2022information},
under the more demanding multinomial channel, yet under the restrictive
constraint on $\beta\in(\frac{1}{2\log|{\cal A}|},\frac{1}{\log|{\cal A}|})$,
which excludes very short molecules. 
\begin{IEEEproof}
For the DNA storage channel, the number of unique objects is the number
of unique molecules of length $L_{K}=\beta\log K$, given by $n\equiv|{\cal A}|^{L_{K}}=K^{\beta\log|{\cal A}|}$,
and the total number of objects is $ng_{n}\equiv K$, that is, $g_{n}=K^{1-\beta\log|{\cal A}|}$,
whereas $N_{K}=r_{n}$. The converse bound of Theorem \ref{thm: noiseless kernel bounds}
then implies that 
\[
\frac{\log M_{\text{DNA}}^{*}(L_{K},V_{L_{K}},N_{K},\epsilon_{K})}{K^{\beta\log|{\cal A}|}}\leq\frac{1}{2}\log\left[N_{K}\wedge(eK^{1-\beta\log|{\cal A}|})\right]+o_{K}(1).
\]
If, e.g., $N_{K}=K^{1-\beta\log|{\cal A}|}$ (that is, $r_{n}=g_{n})$
we get 
\[
\frac{\log M_{\text{DNA}}^{*}(L_{K},V_{L_{K}},N_{K},\epsilon_{K})}{LK^{\beta\log|{\cal A}|}}\leq\frac{1-\beta\log|{\cal A}|}{2\beta}+o\left(\frac{1}{\log K}\right).
\]
The achievability bound of Theorem \ref{thm: noiseless kernel bounds}
requires the condition $n=\Omega(g_{n}^{1+\zeta})$, which translates
into $\beta>\frac{1}{2\log|{\cal A}|}$, and then implies that 
\[
\frac{\log M_{\text{DNA}}^{*}(L_{K},V_{L_{K}},N_{K},\epsilon_{K})}{LK^{\beta\log|{\cal A}|}}\geq\frac{1}{2\beta}\frac{\log(N_{K})}{\log K}-\Psi\left(\frac{N_{K}}{K^{1-\beta\log|{\cal A}|}}\right)+o\left(\frac{1}{\log K}\right).
\]
Using $N_{K}=K^{1-\beta\log|{\cal A}|}$ results 
\begin{align}
\frac{\log M_{\text{DNA}}^{*}(L_{K},V_{L_{K}},N_{K},\epsilon_{K})}{LK^{\beta\log|{\cal A}|}} & \geq\frac{1}{2\beta}\frac{\log(K^{1-\beta\log|{\cal A}|})}{\log K}-\frac{\Psi(1)}{\beta\log K}+o\left(\frac{1}{\log K}\right)\\
 & =\frac{1-\beta\log|{\cal A}|}{2\beta}-\frac{2.773}{2\beta}\cdot\frac{1}{\log K}+o\left(\frac{1}{\log K}\right).\label{eq: explicity lower bound DNA storage}
\end{align}
Thus, in this regime for $\beta$ the converse and achievable bounds
only differ in $O(\log^{-1}K)$ term. It should be mentioned that
using the approximately optimized value of $N_{K}=0.4K^{1-\beta\log|{\cal A}|}$
slightly improves the factor of $1/\log K$ from $\frac{2.773}{2\beta}$
to $\frac{2.59}{2\beta}$.
\end{IEEEproof}
\begin{example}
Consider a DNA storage system with $|{\cal A}|=4$. We compare the
number of achievable bits, that is, the asymptotic lower bound on
$\log M_{\text{DNA}}^{*}(L_{K},V_{L_{K}},N_{K},\epsilon_{K})$ of
(\ref{eq: explicity lower bound DNA storage}) (without the $o(\log^{-1}K)$
term) to the total number of nucleotides, in the short-molecule regime.
As mentioned in \cite[Sec. 1]{shomorony2022information}, just $5[\text{grams}]$
of DNA contain about $KL=4\cdot10^{21}$ nucleotides. Thus, e.g.,
if $\beta=\frac{0.76}{\log(4)}$ then Fig. \ref{fig: log cardinality lower bound versus KL}
shows that these $5[\text{grams}]$ store over $1.253\cdot10^{16}[\text{nats}]=1.8\cdot10^{16}[\text{bits}]$,
while $L=\beta\W(\frac{KL}{\beta})\approx26$, where here $\W$ is
the Lambert W function. This is a huge amount of stored data, while
the molecule length is rather short and thus amenable for efficient
synthesis and sequencing.
\end{example}

\paragraph*{The assumptions of the theorem}
\begin{enumerate}
\item A simple application of the data-processing theorem implies that the
converse bound is valid for any Markov kernel $W_{n}$, not just noiseless.
\item The condition $n=\Omega(g_{n}^{1+\zeta})$ can be relaxed to just
$n=\omega(g_{n})$. We have used the polynomial factor $\frac{n}{g_{n}}=\Omega(g_{n}^{\zeta})$
for simplicity of exposition. 
\end{enumerate}

\subsection{Proof Outline of the Converse Bound of Theorem \ref{thm: noiseless kernel bounds}}

The proof follows the standard Fano's inequality, which requires bounding
$I(X^{n};Y^{n})$. The frequency-based channel from $X^{n}$ to $Y^{n}$
is a multinomial channel $Y^{n}\sim\Mul(nr_{n},\hat{x}^{n})$, which
is not a memoryless channel, and so a direct analysis of the mutual
information is difficult. Nonetheless, as is well known, the multinomial
distribution can be converted into a Poisson distribution (see Appendix
\ref{sec:The-Poisson-Distribution}). The Poisson distribution is
memoryless, and so the resulting mutual information is amenable for
evaluation. Specifically, we consider $Z^{n}$ to be the output of
a memoryless Poisson channel, with the same input $X^{n}$, and then
relate $I(X^{n};Y^{n})$ to $I(X^{n};Z^{n})$. An optimal input distribution
for the Poisson channel is memoryless $P_{X^{n}}=P_{X}^{\otimes n}$,
which, in turn, requires bounding a single-letter $I(X;Z)$. This
is bounded using the known bounds on the average-power-constrained
Poisson channel \cite[Thm. 7]{lapidoth2008capacity}, in the asymptotic
regime of high power. This results in the term $\frac{1}{2}\log(r_{n})+o_{n}(1)$
in the upper bound. Next, we also note that $I(X^{n};Y^{n})\leq H(X^{n})$,
and that since $X^{n}$ is non-negative integer-valued and $\sum_{i=1}^{n}X_{i}\leq ng_{n}$,
this puts an immediate constraint on the cardinality of the alphabet
of $X^{n}$, and thus on its entropy. Stirling's bound then shows
that it is bounded as $\frac{n}{2}\log(eg_{n})$, which results the
term $\frac{1}{2}\log(eg_{n})+o_{n}(1)$ in the upper bound.

\subsection{Steps of the Proof of the Achievability Bound of Theorem \ref{thm: noiseless kernel bounds}}

The proof of achievability Theorem \ref{thm: noiseless kernel bounds}
is based on the three propositions that will be described next. Here
we will state these propositions, and briefly outline their proofs.
The detailed proof will appear in Sec. \ref{sec: Proof achievability}.
The proof is based on Feinstein\textquoteright s maximal coding bound
\cite{feinstein1954new}, which bounds the maximal error probability
of the optimal codebook of a given cardinality via the cumulative
distribution function (CDF) of the \emph{information density} of the
channel (i.e., the \emph{information spectrum}). Concretely, we use
the extended version stated in \cite[Thm. 20.7]{polyanskiy2023information},
which also takes into account input constraints. Let $P_{Y^{n}\mid X^{n}}$
denote the Markov kernel from the input $X^{n}$ to the output $Y^{n}$,
for which $Y^{n}\mid X^{n}=x^{n}\sim\Mul(nr_{n},\hat{x}^{n})$. Let
$P_{X^{n}}$ denote the input distribution, and $P_{Y^{n}}$ be the
corresponding output distribution. Let the information density be
\begin{equation}
i(x^{n};y^{n}):=\log\frac{P_{Y^{n}\mid X^{n}}(y^{n}\mid x^{n})}{P_{Y^{n}}(y^{n})}.\label{eq: information density}
\end{equation}
The extended Feinstein bound assures the following: For any $\gamma>0$
and $M\in\mathbb{N}_{+}$ there exists a code ${\cal C}_{M}=\{x^{n}(1),\ldots,x^{n}(M)\}$
such that $x^{n}(j)\in F_{n}$ for all $j\in[M]$, and whose maximal
error probability is $\epsilon_{n}$, where
\begin{equation}
\epsilon_{n}P_{X^{n}}(F_{n})\leq\P\left[i(X^{n};Y^{n})\leq\log\gamma\right]+\frac{M}{\gamma}.\label{eq: extended Feinstein's bound}
\end{equation}
Here too, the fact that the channel from $X^{n}$ to $Y^{n}$ is not
memoryless makes a direct analysis of the information spectrum challenging,
and similarly, it is altered to a memoryless Poisson distribution.
To obtain a useful bound, however, it is required to restrict the
input distribution to a finite support. The result is that we show
that there exists a codebook whose number of codewords is roughly
$M\approx e^{nI(X;Z)}$, where $I(X;Z)$ is the mutual information
of the Poisson channel, and the error probability upper bounded, by
a bound which can be made vanishing. This first step is summarized
in the following proposition. 
\begin{prop}
\label{prop: Feinstein's bound for Poisson}Let $P_{X}$ be a distribution
such that $\supp(P_{X})\subseteq[s_{n}]=\{1,2,\ldots,s_{n}\}$ for
some $s_{n}\in\mathbb{N}_{+}$. Let $F_{n}:=\{x^{n}\in\mathbb{N}^{n}\colon\frac{1}{n}\sum_{i=1}^{n}x_{i}=g_{n}\}$
be a set of input vectors. Also let $Z\mid X=x\sim\Poi(\frac{r_{n}}{g_{n}}x)$,
and let $\delta_{n}\in(0,\frac{r_{n}}{g_{n}}s_{n})$, where $\frac{r_{n}}{g_{n}}s_{n}\geq12\pi e^{2}$.
Then, there exists a code ${\cal C}_{M}\subset F_{n}$ of $M$ codewords
with 
\[
\log M=nI(X;Z)-3n\delta_{n}-\frac{1}{2}\log(6\pi nr_{n}),
\]
whose maximal error probability $\epsilon_{n}$ on the multinomial
channel from $X^{n}$ to $Y^{n}$ is bounded as 
\begin{equation}
\epsilon_{n}\leq\frac{11}{P_{X}^{\otimes n}(F_{n})}\left[\sqrt{nr_{n}}\exp\left[-n\delta_{n}^{2}\cdot\left(\frac{2}{\log^{2}\frac{r_{n}s_{n}}{g_{n}}}\wedge\frac{g_{n}}{19r_{n}s_{n}\log^{2}s_{n}}\right)\right]+e^{-n\delta_{n}}\right].\label{eq: Poissonized Feinstein bound}
\end{equation}
\end{prop}
\begin{IEEEproof}[Proof outline of Prop. \ref{prop: Feinstein's bound for Poisson}]
Feinstein's bound is based on the information spectrum $\P[i(X^{n};Y^{n})\leq\log\gamma]$
for $\gamma>0$. In order to relate this to the information spectrum
of a memoryless Poisson channel $(X^{n},Z^{n})$ we first relate the
information density $i(x^{n};y^{n})$ to that of Poisson, i.e., to
\[
\tilde{i}(x^{n};z^{n}):=\log\frac{P_{Z^{n}\mid X^{n}}(z^{n}\mid x^{n})}{P_{Z^{n}}(z^{n})}.
\]
We show that the modification of the information density leads to
an additive loss term in $\log M$ given by $\frac{1}{2}\log(6\pi nr_{n})$,
which will be negligible after normalizing by $n$. Second, we replace
the randomness over $(X^{n},Y^{n})$ in the information spectrum with
that of $(X^{n},Z^{n})$, using the Poissonization of the multinomial
effect (see Fact \ref{fact: Poissonization of the multinomial distribution}).
As a result, the analysis of the information spectrum of the channel
from $X^{n}$ to $Y^{n}$ is altered to the analysis of the information
spectrum $\P[\tilde{i}(X^{n};Z^{n})\leq\log\gamma]$. Since the Poisson
channel is memoryless, if we further restrict $P_{X^{n}}$ to a product
distribution $P_{X}^{\otimes n}$, then $\tilde{i}(X^{n};Z^{n})$
is a sum of IID RVs, for which tail bounds can be readily derived.
Before discussing the derivation of this bound, we highlight that
the required bound should decay faster than its decay for standard
analysis of memoryless channels. In the standard analysis, $\gamma$
is chosen so that $\P[\tilde{i}(X^{n};Z^{n})\leq\log\gamma]\to0$
as $n\to\infty$, albeit with an arbitrary slow rate (e.g., in the
proof of \cite[Thm. 19.8]{polyanskiy2023information}). Here, this
probability is multiplied by a term that scales as $\Theta(\frac{\sqrt{nr_{n}}}{P_{X}^{\otimes n}(F_{n})})$,
and so obtaining a vanishing upper bound on $\epsilon_{n}$ requires
a bound which is $o(\frac{P_{X}^{\otimes n}(F_{n})}{\sqrt{nr_{n}}})$.
To obtain the desired upper bound, we separate the analysis of the
randomness of $Z^{n}$ conditioned on $X^{n}=x^{n}$ from the randomness
of $X^{n}$. To analyze the randomness over $Z^{n}$ conditioned on
$X^{n}=x^{n}$, we note that $\tilde{i}(x^{n};Z^{n})$ is a function
of $n$ independent Poisson RVs. We show that under the restricted
support assumption, that is, $\supp(P_{X})\subseteq[s_{n}]$, it holds
that $\tilde{i}(x^{n};Z^{n})$ is a Lipschitz function with semi-norm
$\log s_{n}$. In turn, this allows us to use the concentration bound
of Lipschitz functions of Poisson RVs due to Bobkov and Ledoux \cite[Prop. 11]{bobkov1998modified}
(see Appendix \ref{sec:Poisson-concentration-of} for a brief overview).
Thus we show that $\tilde{i}(x^{n};Z^{n})$ is close to its expected
value, denoted as $J(x^{n}):=\sum_{i=1}^{n}\E[\log P_{Z\mid X}(Z_{i}\mid x_{i})\mid X_{i}=x_{i}]=\sum_{i=1}^{n}J(x_{i})$.\footnote{With a slight abuse of notation, we use $J(\cdot)$ for both scalar
and vector inputs, with the common convention that the value of vector
inputs is the sum of the value of the scalar function of the coordinates. } Hence, under the choice of memoryless input distribution, $J(x^{n})$
is also a sum of independent RVs. We then prove that $J(x)\in[-\log\frac{r_{n}s_{n}}{g_{n}},0]$,
that is, $J(X^{n})$ is a sum of \emph{bounded} RVs. An application
of Hoeffding's inequality then shows that $J(X^{n})$ concentrates
to its expected value $-H(Z\mid X)$. Combining the concentration
results of both $J(X^{n})$ and $\tilde{i}(x^{n};Z^{n})$ leads to
an upper bound on $\P[\tilde{i}(X^{n};Z^{n})\leq\log\gamma]$, then
to an upper bound on $\P[i(X^{n};Y^{n})\leq\log\gamma]$, and finally,
to the claimed upper bound on the error probability $\epsilon_{n}$,
via Feinstein's bound. 
\end{IEEEproof}
Further evaluation of the Feinstein-based bound (\ref{eq: Poissonized Feinstein bound})
in Prop. \ref{prop: Feinstein's bound for Poisson} requires two tasks:
First, evaluating the mutual information $I(X;Z)$ over the Poisson
channel, and second, evaluating the probability that a randomly chosen
codeword meets the constraint, that is $P_{X}^{\otimes n}(F_{n})$.
This is the content of the next two propositions, beginning with the
former.

Let $\tilde{X}$ be a continuous input RV. As mentioned, under the
average-power input constraint $\E[\tilde{X}]\leq g_{n}$, the optimal
input distribution for a Poisson channel is $\tilde{X}\sim\Gam(\frac{1}{2},2g_{n})$
\cite{martinez2007spectral,lapidoth2008capacity}. This is a continuous
distribution supported on $\mathbb{R}_{+}$, and thus unsuitable to
the frequency-based channel, which accepts non-negative \emph{integer}
inputs. Furthermore, the bound of Prop. \ref{prop: Feinstein's bound for Poisson}
is based on the assumption that $\supp(P_{X})\subseteq[s_{n}]$, where
$s_{n}\in\mathbb{N}_{+}$ is finite. In order to obtain a valid lower
bound on the mutual information, we modify the gamma distribution
of $\tilde{X}$ by first truncating (or restricting) it to a judicious
choice of interval ${\cal S}_{n}$, and then rounding it to be integer
valued so that the resulting RV is supported on $[s_{n}]$. We will
use the following definition for truncation:
\begin{defn}
\label{def: truncation}Let $A$ be a real RV, and let ${\cal S}\subset\mathbb{R}$
be such that $\P[A\in{\cal S}]>0$. The truncation of $A$ to a support
${\cal S}$ is the RV $A_{\mid{\cal S}}$ which satisfies that for
any Borel set ${\cal A}\in\mathfrak{B}(\mathbb{R})$,
\begin{equation}
\P\left[A_{\vert{\cal S}}\in{\cal A}\right]=\frac{\P\left[A_{\vert{\cal S}}\in{\cal A}\cap{\cal S}\right]}{\P\left[A\in{\cal S}\right]}.\label{eq: truncation to a given support}
\end{equation}
\end{defn}
\begin{prop}
\label{prop: Truncated gamma MI}Assume that $g_{n}\to\infty$, and
that $\underline{c}g_{n}\leq r_{n}\leq eg_{n}$ for some $\underline{c}\in(0,e)$.
Let $\rho\in(0,1)$ be given, and consider the interval 
\begin{equation}
{\cal S}_{n}=\left[\frac{1}{g_{n}^{1+3\rho}},g_{n}^{1+\rho}\right]\in\mathbb{R}_{+}.\label{eq: truncation interval}
\end{equation}
Let $\tilde{X}\sim\Gam(\frac{1}{2},2g_{n})$, and let $X=\lceil\tilde{X}_{\vert{\cal S}_{n}}\rceil$,
i.e., $\tilde{X}$ is first truncated to ${\cal S}_{n}$ and then
rounded upward to the nearest integer. Further let $Z\mid X=x\sim\Poi(\frac{r_{n}}{g_{n}}x)$
for $x\in\mathbb{R}_{+}$. Then, there exists $n_{0}$ (which depends
on $(\underline{c},\rho)$ and $\{g_{n}\}$), such that for all $n\geq n_{0}$\textbf{
\begin{equation}
I(X;Z)\geq\frac{1}{2}\log r_{n}-\Psi\left(\frac{r_{n}}{g_{n}}\right)-o_{n}(1).\label{eq: lower bound on Poisson MI for truncated and rounded}
\end{equation}
}
\end{prop}
\begin{IEEEproof}[Proof outline of Prop. \ref{prop: Truncated gamma MI}]
The proof bounds the loss in the achievable mutual information $I(X;Z)$
when the asymptotically ideal $\tilde{X}\sim\Gam(\frac{1}{2},2g_{n})$
is truncated to ${\cal S}_{n}$ and then upward rounded to an integer.
We begin by analyzing $\overline{X}:=\tilde{X}_{\mid{\cal S}_{n}}$.
A direct analysis of the reduction in mutual information when modifying
$\tilde{X}$ to $\overline{X}$ appears to be cumbersome, and we thus
take an indirect route, which exploits the relation between mutual
information and optimal estimation over the Poisson channel \cite{guo2008mutual,atar2012mutual,jiao2013pointwise,wang2014bregman,dytso2020estimation,dytso2020vector,dytso2021properties},
\cite[Ch. 8]{guo2013interplay}, which we next briefly review. Let
$\ell(u,v)\equiv\ell_{\text{Poi}}(u,v):=v-u+u\log\frac{u}{v}$ be
the Poisson error function. For a positive random variable $U$, we
let $V\mid U=u\sim\Poi(u)$. Let $\hat{U}$ be an estimator of $U$
based on $V$. Then, since $\ell(u,v)$ is the Bregman divergence
\cite{bregman1967relaxation} associated with the Poisson distribution,
it holds that the minimal estimation error is obtained by the expected
mean $\E[U\mid V]$ and the \emph{minimum mean Poisson error} (MMPE)
is 
\begin{align}
\mmpe(U) & =\min_{\hat{U}}\E\left[\ell(U,\hat{U})\right]\\
 & =\E\left[\ell\left(U,\E[U\mid V]\right)\right]\\
 & =\E\left[U\log\frac{U}{\E[U\mid V]}\right].
\end{align}
The following relation between the MMPE and the mutual information
was established in \cite[Corollary 1]{guo2008mutual}:
\begin{thm}[{I-MMPE relation \cite[Corollary 1]{guo2008mutual}}]
\label{thm:I-MMPE}Assume that $\E[U\log U]<\infty$ and let $V_{a}\mid U=u\sim\Poi(au)$
for $a>0$. Then, 
\[
I(U;V_{\gamma})=\int_{0}^{\gamma}\mmpe(aU)\cdot\frac{\d a}{a}.
\]
\end{thm}
Following a similar analysis for the Gaussian channel \cite[Lemma 2]{wu2011functional},
we analyze the difference between $\mmpe(aU)$ and $\mmpe(a\overline{U})$,
where $\overline{U}$ is a truncated version of $U$ (Lemma \ref{lem:truncation in the MI general}).
We show that this difference is controlled by three terms 
\[
\gamma\cdot s_{\text{max}}\cdot\P\left[U\in[0,s_{\text{min}})\right]+\gamma\E\left[U\log U\cdot\I\left\{ U\in(s_{\max},\infty)\right\} \right]+\gamma\E\left[U\log\frac{1}{s_{\text{min}}}\cdot\I\left\{ U\in(s_{\max},\infty)\right\} \right]
\]
where ${\cal S}:=[s_{\text{min}},s_{\text{max}}]$ is the set used
for truncation. Accordingly, we show that this difference is small
for the specific gamma distribution of $\tilde{X}$ and the truncation
set of interest (Lemma \ref{lem: MI loss due to trunaction terms})
when $U$ follows the distribution of $\tilde{X}\sim\Gam(\frac{1}{2},2g_{n})$.
The difference between the MMPE of $\tilde{X}$ and $\overline{X}$
is then translated, via the I-MMPE relation in Theorem \ref{thm:I-MMPE},
to a difference between their mutual information, which is eventually
shown to be a negligible $o_{n}(1)$ term in the regime of interest.
Thus, there is no essential loss in mutual information due to the
truncation of the gamma distribution.

Next, we consider the influence of rounding $\overline{X}$ to the
integer $X=\lceil\overline{X}\rceil$. In contrast to truncation,
it appears that the rounding operation leads to a loss in the mutual
information, and we upper bound this loss as $\Psi(\frac{r_{n}}{g_{n}})$
(Lemma \ref{lem: mutual information of  rounded gamma}).\textbf{
}This is the source of the additive loss term $\Psi(\frac{r_{n}}{g_{n}})$
that appears in the statement of the proposition. Specifically, from
$I(X;Z)=H(Z)-H(Z\mid X)$ and $I(\overline{X};\overline{Z})=H(\overline{Z})-H(\overline{Z}\mid\overline{X})$
with $\overline{Z}\mid\overline{X}=\overline{x}\sim\Poi(\frac{r_{n}}{g_{n}}\overline{x})$,
we may compare the mutual information values by separately comparing
the conditional entropy values and the output entropy values.\textbf{
}First, we use properties of the entropy of the Poisson PMF and the
input gamma distribution to show that $H(\overline{Z}\mid\overline{X})\leq H(Z\mid X)+o_{n}(1)$.
Second, we show that $H(\overline{Z})$ is only larger than $H(Z)$
by at most $\Psi(\frac{r_{n}}{g_{n}})$.\textbf{ }The proof of this
result relies on the infinite divisibility property of the Poisson
distribution. By writing $X=\overline{X}+D$, where $D\in[0,1]$,
we may also write $Z=\overline{Z}+\acute{Z}$ where $\acute{Z}\mid D=d\sim\Poi(\frac{r_{n}}{g_{n}}d)$
(note, however, that $Z$ are $\acute{Z}$ are not independent). We
then use the bounding method used, e.g., in \cite[Prop. 8]{polyanskiy2016wasserstein}.
We relate $H(\overline{Z})-H(Z)$ to the \emph{maximum entropy }that
is possible for a non-negative integer-valued RV whose expectation
is less than $\E[Z-\overline{Z}]$. This maximum entropy is well-known
to be the entropy of a proper geometric RV (Lemma \ref{lem:Geometric max entropy under mean constraint}),
given by the function $\Psi(\mu)$, assuming that the allowed mean
is $\mu$. Combining the comparison between the values of the conditional
entropy and the values of the output entropy leads to the required
comparison between the mutual information values. The final bound
is obtained by combining the effects of both the truncation and the
rounding. 
\end{IEEEproof}
Following the analysis of $I(X;Z)$ for a proper choice of $X$, it
remains to bound $P_{X^{n}}(F_{n})$. Since the term multiplying $P_{X^{n}}(F_{n})$
in (\ref{eq: Poissonized Feinstein bound}) decays super-polynomially
with $n$, the error probability $\epsilon_{n}$ may decay to zero,
even if $P_{X^{n}}(F_{n})$ decays to zero, although polynomially. 
\begin{prop}
\label{prop: Constant weight input high probability}Let $\zeta>0$
and $\rho\in(0,\frac{1}{4}\vee\frac{2\zeta}{3})$ be given. For $i\in[n]$,
let $\{\tilde{X}_{i}\}_{i\in[n]}$ be IID with $\tilde{X}_{i}\sim\Gam(\frac{1}{2},2g_{n})$,
and let $X_{i}=\lceil\tilde{X}_{\vert{\cal S}_{n}}\rceil$ where ${\cal S}_{n}=[g_{n}^{-(1+3\rho)},g_{n}^{1+\rho}]$.
Let $F_{n}(\tau):=\{x^{n}\in\mathbb{N}^{n}\colon\frac{1}{n}\sum_{i=1}^{n}x_{i}=\tau\}$.
Assume that $n=\Omega(g_{n}^{1+\zeta})$. Then, there exists a sequence
$\varsigma_{n}=o_{n}(1)$ and $\tau_{n}\in[g_{n}(1+\varsigma_{n})]$
such that 
\[
P_{X^{n}}\left[F_{n}(\tau_{n})\right]\geq\frac{1}{3ng_{n}}
\]
for all $n$ sufficiently large. 
\end{prop}
\begin{IEEEproof}[Proof outline of Prop. \ref{prop: Constant weight input high probability}]
The main technical aspect of the proof is to find a right-tail bound
on $\sum_{i=1}^{n}\overline{X}_{i}$, where $\{\overline{X}_{i}\}_{i\in[n]}$
are IID, and each is distributed according to the $\Gam(\frac{1}{2},2g_{n})$
distribution \emph{truncated} to ${\cal S}_{n}=[g_{n}^{-(1+3\rho)},g_{n}^{1+\rho}]$.
The gist of the proof is to define a proper generative model for $\{\overline{X}_{i}\}_{i\in[n]}$.
To this end, we define $\{\tilde{X}_{i,j}\}_{i\in[n],j\in\mathbb{N}_{+}}$
to be a double-index array of IID RV, distributed according to $\tilde{X}_{i,j}\sim\Gam(\frac{1}{2},2g_{n})$.
We then define $\overline{X}_{i}=\tilde{X}_{J^{*}(i)}$ where $J^{*}(i)\in\mathbb{N}_{+}$
is the minimal index such that $\tilde{X}_{i,j}\in{\cal S}_{n}$.
Evidently, $\{\overline{X}_{i}\}_{i\in[n]}$ are IID, and have the
required truncated gamma distribution. Now, using the fact that ${\cal S}_{n}$
is a high probability set, we show that, with high probability, $\tilde{X}_{i,1}$
is in ${\cal S}_{n}$ for most indices, and so $\overline{X}_{i}=\tilde{X}_{1}$
for most indices, which we denote by $n-\ell$. For the remaining
$\ell$ indices, it holds that $\overline{X}_{i}\leq g_{n}^{1+\rho}$,
and so their effect on $\frac{1}{n}\sum_{i=1}^{n}\overline{X}_{i}$
is controlled. Consequently, loosely speaking, $\sum_{i=1}^{n}\overline{X}_{i}$
is dominated by $\sum_{i=1}^{n}\tilde{X}_{i,1}$, up to terms which
are eventually negligible. Taking into account the number of possibilities
for $\{J^{*}(i)\}_{i\in[n]}$ (under the high probability event),
and using standard tail bounds on the gamma distribution, we show
that $\sum_{i=1}^{n}\overline{X}_{i}\leq ng_{n}(1+\overline{\varsigma}_{n})$
with probability larger than, say, $1/2$. In turn, this is also true
for the upward rounded $X_{i}=\lceil\overline{X}_{i}\rceil$ with
$\overline{\varsigma}_{n}$ replaced by $\varsigma_{n}$ which is
essentially the same. Since $\sum_{i=1}^{n}X_{i}$ is an integer,
roughly upper bounded by $ng_{n}$, there must exists $n\tau_{n}\in\mathbb{N}_{+}$,
with $\tau_{n}\leq g_{n}(1+\varsigma_{n})$ such that the probability
of $F_{n}(\tau_{n})$, as defined in the proposition, is $\Omega(\frac{1}{ng_{n}})$,
as claimed.
\end{IEEEproof}
The proof of the achievability bound of Theorem \ref{thm: noiseless kernel bounds}
then directly combines the above propositions: First, Prop. \ref{prop: Feinstein's bound for Poisson}
leads to a Feinstein-based bound on the number of codewords $M$ and
error probability $\epsilon_{n}$, which depends on $I(X;Z)$ of the
Poisson channel with integer inputs, and the probability of an input
vector with a fixed sum $\tau_{n}$ below $ng_{n}$. Prop. \ref{prop: Truncated gamma MI}
lower bounds $I(X;Z)$, and Prop. \ref{prop: Constant weight input high probability}
bounds $P_{X^{n}}[F_{n}(\tau_{n})]$. Analyzing the leading terms
in the resulting mutual information and error probability leads to
the claimed result. 

\section{Proof of Theorem \ref{thm: noiseless kernel bounds}\label{sec:Proofs}}

\subsection{Proof of the converse bound of Theorem \ref{thm: noiseless kernel bounds}}
\begin{IEEEproof}[Proof of the converse bound of Theorem \ref{thm: noiseless kernel bounds}]
Let ${\cal C}_{M}$ be a code of cardinality $M$, whose maximal
error probability is $\epsilon_{n}$, and for which each codeword
$x^{n}(j)\in{\cal C}_{M}$ satisfies $\sum_{i=1}^{n}x_{i}(j)=n\underline{g}_{n}$
for some $\underline{g}_{n}\leq g_{n}$. By Fano's inequality (e.g.
\cite[Thm. 20.6]{polyanskiy2023information}) it holds that 
\begin{equation}
\log M\leq\frac{1}{1-\epsilon_{n}}\left(\hbin(\epsilon_{n})+\sup_{P_{X^{n}}\colon\supp(P_{X^{n}})\subset\mathbb{N}^{n},\;P_{X^{n}}\left(\frac{1}{n}\sum_{i=1}^{n}X_{i}=\underline{g}_{n}\right)=1}I(X^{n};Y^{n})\right)\label{eq: Fano inequality}
\end{equation}
where $Y^{n}\sim\Mul(nr_{n},\frac{1}{n\underline{g}_{n}}X^{n})$.
We evaluate this bound by further analyzing $I(X^{n};Y^{n})$. Set
$\eta\in(0,1)$, and let $Z^{n}=(Z_{1},\ldots,Z_{n})$ be a vector
of independent components, such that $Z_{i}\mid X_{i}=x_{i}\sim\Poi(\frac{1}{1-\eta}\cdot\frac{r_{n}}{\underline{g}_{n}}x_{i})$.
We next relate $I(X^{n};Z^{n})$ to $I(X^{n};Y^{n})$. Let $Q:=\sum_{i=1}^{n}Z_{i}\sim\Poi(\frac{1}{1-\eta}nr_{n})$
be the random number of output objects in the Poisson model. Let $\{S_{i}\}_{i=1}^{\infty}$
be drawn as in the problem formulation (Sec. \ref{subsec:The-Concentration-Input-Channel}),
and with a slight abuse of notation, let $Y^{n}(q)\sim\Mul(q,\frac{1}{\underline{g}_{n}}X^{n})$
be a sequence of RVs, with the coupling that $Y^{n}(q)$ is the histogram
of $S^{q}$. Thus, $Y^{n}\eqd Y^{n}(nr_{n})$. Furthermore, since
$\P[S_{1},S_{2},\ldots,S_{q}\mid X^{n}=x^{n}]=\P[S_{\pi(1)},S_{\pi(2)},\ldots,S_{\pi(q)}\mid X^{n}=x^{n}]$
for any permutation $\pi$ in the symmetric group of degree $q$,
it holds that $I(X^{n};S^{q})=I(X^{n};Y^{n}(q))$. The data-processing
inequality then implies that for any $q_{1}\leq q_{2}$ 
\begin{equation}
I(X^{n};Y^{n}(q_{1}))=I(X^{n};S^{q_{1}})\leq I(X^{n};S^{q_{2}})=I(X^{n};Y^{n}(q_{2})).\label{eq: monotonicity of MI multinomial}
\end{equation}
Now, from the Poissonization of the multinomial distribution effect
(Fact \ref{fact: Poissonization of the multinomial distribution})
it holds that $Z^{n}\mid X^{n},Q=q\eqd Y^{n}(q)\mid X^{n}$ for any
$q\in\mathbb{N}$. Hence, 
\begin{align}
I(X^{n};Z^{n}) & =I(X^{n};Z^{n},Q)\\
 & =I(X^{n};Q)+I(X^{n};Z^{n}\mid Q)\\
 & \geq I(X^{n};Z^{n}\mid Q)\\
 & =\sum_{q=0}^{\infty}\P[Q=q]\cdot I(X^{n};Y^{n}(q))\\
 & \geq\sum_{q=nr_{n}}^{\infty}\P[Q=q]\cdot I(X^{n};Y^{n}(q))\\
 & \trre[\geq,a]\P[Q\geq nr_{n}]\cdot I(X^{n};Y^{n})\\
 & \trre[\geq,b]\left(1-e^{-\frac{\eta^{2}}{2(1-\eta)}nr_{n}}\right)\cdot I(X^{n};Y^{n}),\label{eq: relating Poisson and multinomial capacity}
\end{align}
where $(a)$ follows from the monotonicity property in (\ref{eq: monotonicity of MI multinomial}),
and $(b)$ from Chernoff's bound for Poisson RVs (Lemma \ref{lem: Poisson chernoff}),
which implies that $\P[Q\geq nr_{n}]\geq1-\exp[-\frac{\eta^{2}}{2(1-\eta)}nr_{n}]$
for any $\eta\in(0,1)$. Therefore, 
\begin{align}
 & \sup_{P_{X^{n}}\colon\supp(P_{X^{n}})\subset\mathbb{N}^{n},\;P_{X^{n}}\left(\frac{1}{n}\sum_{i=1}^{n}X_{i}=\underline{g}_{n}\right)=1}I(X^{n};Y^{n})\nonumber \\
 & \leq\sup_{P_{X^{n}}\colon\supp(P_{X^{n}})\subset\mathbb{N}^{n},\;\frac{1}{n}\sum_{i=1}^{n}\E[X_{i}]=\underline{g}_{n}}I(X^{n};Y^{n})\\
 & \leq\sup_{P_{X^{n}}\colon\supp(P_{X^{n}})\subset\mathbb{N}^{n},\;\frac{1}{n}\sum_{i=1}^{n}\E[X_{i}]\leq\underline{g}_{n}}I(X^{n};Y^{n})\\
 & \trre[\leq,a]\frac{1}{1-e^{-\frac{\eta^{2}}{2(1-\eta)}nr_{n}}}\sup_{P_{X^{n}}\colon\supp(P_{X^{n}})\subset\mathbb{R}^{n},\;\frac{1}{n}\sum_{i=1}^{n}\E[X_{i}]\leq\underline{g}_{n}}I(X^{n};Z^{n})\\
 & \trre[=,b]\frac{1}{1-e^{-\frac{\eta^{2}}{2(1-\eta)}nr_{n}}}\sup_{P_{X^{n}}\colon\supp(P_{X^{n}})\subset\mathbb{R}^{n},\;\frac{1}{n}\sum_{i=1}^{n}\E[X_{i}]\leq\frac{r_{n}}{1-\eta}}I(X^{n};\hat{Z}^{n})\\
 & =\frac{1}{1-e^{-\frac{\eta^{2}}{2(1-\eta)}nr_{n}}}n\cdot\sup_{P_{X}\colon\supp(P_{X})\subset\mathbb{R}^{n},\;\E[X]\leq\frac{r_{n}}{1-\eta}}I(X;\hat{Z})\\
 & \trre[=,c]n\left(\frac{1}{1-e^{-\frac{\eta^{2}}{2(1-\eta)}nr_{n}}}\right)\cdot\left[\frac{1}{2}\log\left(\frac{r_{n}}{1-\eta}\right)+o_{r_{n}}(1)\right]\\
 & =n\left(\frac{1}{1-e^{-\frac{\eta^{2}}{2(1-\eta)}nr_{n}}}\right)\cdot\left[\frac{1}{2}\log(r_{n})-\frac{1}{2}\log(1-\eta)+o_{r_{n}}(1)\right]\\
 & =n\cdot\left[\frac{1}{2}\log(r_{n})+\frac{e^{-\frac{\eta^{2}}{2(1-\eta)}nr_{n}}}{1-e^{-\frac{\eta^{2}}{2(1-\eta)}nr_{n}}}\frac{1}{2}\log r_{n}-\left(\frac{1}{1-e^{-\frac{\eta^{2}}{2(1-\eta)}nr_{n}}}\right)\frac{1}{2}\log(1-\eta)+o_{r_{n}}(1)\right],\label{eq: Poisson capacity upper bound on the MI}
\end{align}
where $(a)$ follows from (\ref{eq: relating Poisson and multinomial capacity}),
in $(b)$ we have defined $\hat{Z}_{n}$ as a Poisson channel with
a unity gain, that is $\hat{Z}_{i}\mid X_{i}=x_{i}\sim\Poi(x_{i})$,
$(c)$ follows from the asymptotic expression of the mean-constrained
Poisson channel capacity in \cite[Thm. 7, Eq. (23)]{lapidoth2008capacity}.
Choosing $\eta\equiv\eta_{n}=(nr_{n})^{1/2-\rho}$ for some $\rho\in(0,1/2)$
shows that if $\epsilon_{n}\to0$ as as $n\to\infty$ then it must
hold that
\[
\log M\leq\frac{1}{2}n\left[\log(r_{n})+o_{n}(1)\right].
\]
Next, it also holds that 
\begin{align}
 & \sup_{P_{X^{n}}\colon\supp(P_{X^{n}})\subset\mathbb{N}^{n},\;P_{X^{n}}\left(\frac{1}{n}\sum_{i=1}^{n}X_{i}=\underline{g}_{n}\right)=1}I(X^{n};Y^{n})\nonumber \\
 & \leq\sup_{P_{X^{n}}\colon\supp(P_{X^{n}})\subset\mathbb{N}^{n},\;P_{X^{n}}\left(\frac{1}{n}\sum_{i=1}^{n}X_{i}=\underline{g}_{n}\right)=1}H(X^{n})\\
 & \leq\log\left|\left\{ x^{n}\subset\mathbb{N}^{n}\colon\;\frac{1}{n}\sum_{i=1}^{n}x_{i}=\underline{g}_{n}\right\} \right|\\
 & \trre[=,a]\log{n\underline{g}_{n}+n-1 \choose n-1}\\
 & \trre[\leq,b]\log{ng_{n}+n-1 \choose n-1}\\
 & \trre[=,c]\log\left[\left(\frac{(ng_{n}+n-1)e}{n-1}\right)^{n-1}\frac{1}{\sqrt{2\pi(n-1)}}\exp\left(-\frac{(n-1)^{2}(1+o(1))}{2(ng_{n}+n-1)}\right)\right]+o(1)\\
 & \leq n\left[\log eg_{n}+o(1)\right],\label{eq: cardinality bound on the MI}
\end{align}
where $(a)$ follows from the stars and bars model, $(b)$ follows
since $\underline{g}_{n}\leq g_{n}$ and the monotonicity of the binomial
coefficient, $(c)$ follows from Stirling's approximation of the binomial
coefficient (see (\ref{eq: binomial approximation small k}) in Fact
\ref{fact: binomial approximtion}, Appendix \ref{sec:Useful-Mathematical-Results}).
Combining both (\ref{eq: Poisson capacity upper bound on the MI})
and (\ref{eq: cardinality bound on the MI}) in Fano's inequality
(\ref{eq: Fano inequality}) results the claimed bound.
\end{IEEEproof}

\subsection{Proof of the Achievability Bound of Theorem \ref{thm: noiseless kernel bounds}
\label{sec: Proof achievability}}

In this section, we prove the achievability bound of Theorem \ref{thm: noiseless kernel bounds}.
To this end, we prove Props. \ref{prop: Feinstein's bound for Poisson},
\ref{prop: Truncated gamma MI} and \ref{prop: Constant weight input high probability}
one after the other, and then combine them in order to complete the
proof of the bound.
\begin{IEEEproof}[Proof of Prop. \ref{prop: Feinstein's bound for Poisson}]
Our goal is to analyze the probability on the right-hand side of
(\ref{eq: extended Feinstein's bound}), which is typically simple
whenever $i(X^{n};Y^{n})$ is a sum of IID RVs. To approach this,
we further choose a scalar distribution $P_{X}$, and, as common,
restrict $P_{X^{n}}$ to be the product distribution $P_{X^{n}}=P_{X}^{\otimes n}$.
However, even under this choice, $i(X^{n};Y^{n})$ is not a sum of
IID RVs, since $P_{Y^{n}\mid X^{n}}$ is not a memoryless channel.
We will transform the analysis of this probability to the analysis
of sum of IID RVs in two steps, first by relating $i(x^{n};y^{n})$
to the information density of a memoryless Poisson channel, and second,
by relating the probability of events under the original channel to
the probability of events under this Poisson channel. Concretely,
let $Z^{n}$ be the output of a channel, such that conditioned on
$X^{n}=x^{n}$ it holds that $Z_{i}\sim\Poi(\frac{r_{n}}{g_{n}}x_{i})$,
and where the components of $Z^{n}$ are independent. Let $P_{Z^{n}\mid X^{n}}$
denote the Markov kernel from the input $X^{n}$ to the output $Z^{n}$.
Recall that $Q=\sum_{i=1}^{n}Z_{i}$, and $\sum_{i=1}^{n}Y_{i}=nr_{n}$
with probability $1$. Then, for any $x^{n}\in\mathbb{N}^{n}$ and
$y^{n}$ with $\sum_{i=1}^{n}y_{i}=nr_{n}$, it holds that
\begin{align}
\frac{P_{Y^{n}\mid X^{n}}(y^{n}\mid x^{n})}{P_{Z^{n}\mid X^{n}}(y^{n}\mid x^{n})} & \trre[=,a]\frac{P_{Z^{n}\mid X^{n},Q}(y^{n}\mid x^{n},nr_{n})}{P_{Z^{n}\mid X^{n}}(y^{n}\mid x^{n})}\\
 & =\frac{P_{Z^{n}\mid X^{n},Q}(y^{n}\mid x^{n},nr_{n})}{\sum_{q=0}^{\infty}P_{Q\mid X^{n}}(q\mid x^{n})\cdot P_{Z^{n}\mid X^{n},Q}(y^{n}\mid x^{n},q)}\\
 & \trre[=,b]\frac{1}{P_{Q\mid X^{n}}(nr_{n}\mid x^{n})}\\
 & \trre[=,c]\frac{(nr_{n})!}{(nr_{n})^{nr_{n}}e^{-nr_{n}}},
\end{align}
where $(a)$ follows from the Poissonization of the multinomial (Fact
\ref{fact: Poissonization of the multinomial distribution}), $(b)$
follows since 
\[
P_{Z^{n}\mid X^{n},Q}(y^{n}\mid x^{n},q)=0
\]
if $q\neq\sum_{i=1}^{n}y_{i}=nr_{n}$, and $(c)$ follows since $Q\mid X^{n}=x^{n}\sim\Poi(nr_{n})$
(i.e., $Q$ is independent of $X^{n}$). Now, Stirling's bound (Fact
\ref{fact: Stirling bound} in Appendix \ref{sec:Useful-Mathematical-Results})
 implies that, with probability $1$
\[
\sqrt{2\pi nr_{n}}\leq\frac{P_{Y^{n}\mid X^{n}}(y^{n}\mid x^{n})}{P_{Z^{n}\mid X^{n}}(y^{n}\mid x^{n})}\leq\sqrt{6\pi nr_{n}}.
\]
Then using $\sum a_{i}/\sum b_{i}\geq\min_{i}(a_{i}/b_{i})$ for reals
$\{(a_{i},b_{i})\}$, we also have
\[
\frac{P_{Z^{n}}(y^{n})}{P_{Y^{n}}(y^{n})}=\frac{\sum_{x^{n}}P_{X^{n}}(x^{n})P_{Z^{n}\mid X^{n}}(y^{n}\mid x^{n})}{\sum_{x^{n}}P_{X^{n}}(x^{n})P_{Y^{n}\mid X^{n}}(y^{n}\mid x^{n})}\geq\frac{1}{\sqrt{6\pi nr_{n}}}.
\]
So, for any $nr_{n}\geq2$,
\begin{align}
 & \P\left[i(X^{n};Y^{n})\leq\log\gamma\right]\nonumber \\
 & =\P\left[\log\frac{P_{Y^{n}\mid X^{n}}(Y^{n}\mid X^{n})}{P_{Y^{n}}(Y^{n})}\leq\log\gamma\right]\\
 & =\P\left[\log\frac{P_{Z^{n}\mid X^{n}}(Y^{n}\mid X^{n})}{P_{Z^{n}}(Y^{n})}+\log\frac{P_{Y^{n}\mid X^{n}}(Y^{n}\mid X^{n})}{P_{Z^{n}\mid X^{n}}(Y^{n}\mid X^{n})}+\log\frac{P_{Z^{n}}(Y^{n})}{P_{Y^{n}}(Y^{n})}\leq\log\gamma\right]\\
 & \trre[\leq,a]\P\left[\log\frac{P_{Z^{n}\mid X^{n}}(Y^{n}\mid X^{n})}{P_{Z^{n}}(Y^{n})}-\frac{1}{2}\log(6\pi nr_{n})\leq\log\gamma\right]\\
 & \trre[\leq,b]e\sqrt{nr_{n}}\P\left[\log\frac{P_{Z^{n}\mid X^{n}}(Z^{n}\mid X^{n})}{P_{Z^{n}}(Z^{n})}-\frac{1}{2}\log(6\pi nr_{n})\leq\log\gamma\right]\\
 & \trre[=,c]e\sqrt{nr_{n}}\P\left[\sum_{i=1}^{n}\log\frac{P_{Z\mid X}(Z_{i}\mid X_{i})}{P_{Z}(Z_{i})}\leq\log\gamma+\frac{1}{2}\log(6\pi nr_{n})\right],\label{eq: Poissonization of the information density CDF}
\end{align}
where $(a)$ follows since $\log\sqrt{2\pi nr_{n}}\geq0$, $(b)$
follows since the probability of any event of a multinomial is upper
bounded, with a proper factor, by the probability of that event under
its Poissonized version \cite[Thm. 5.7 and Corollary 5.9]{mitzenmacher2017probability}
(see Lemma \ref{lem: Poissonization of events} in Appendix \ref{sec:The-Poisson-Distribution}),
and $(c)$ holds since $P_{Z^{n}\mid X^{n}}(Z^{n}\mid X^{n})$ is
a product Markov kernel, which combined with the restriction $P_{X^{n}}=P_{X}^{\otimes n}$
results that $\log\frac{P_{Z^{n}\mid X^{n}}(Z^{n}\mid X^{n})}{P_{Z^{n}}(Z^{n})}$
is decomposed to a sum of IID RVs. 

We continue to upper bound the probability in (\ref{eq: Poissonization of the information density CDF})
over $(X^{n},Z^{n})$. Due to the pre-factor $\Theta(\sqrt{nr}_{n})$,
we will need to show that this probability decays sufficiently fast
in order to obtain a sufficiently strong bound on the probability
in the extended Feinstein bound (\ref{eq: extended Feinstein's bound}).
We again bound in two steps. First, we condition on $X^{n}=x^{n}$,
and analyze the probability with respect to (w.r.t.) the randomness
of $Z^{n}$ and second, we analyze the resulting upper bound w.r.t.
the randomness of $X^{n}$. 

We begin with the first step, for which we recall that the $\supp(P_{X})\subseteq[s_{n}]=\{1,2,\ldots,s_{n}\}$,
and specifically, that $P_{X}(0)=0$. We use this assumption to establish
that for any $x^{n}\in\supp(P_{X}^{\otimes n})$, the RV
\[
f_{x^{n}}(Z^{n}):=\sum_{i=1}^{n}\log\frac{P_{Z\mid X}(Z_{i}\mid x_{i})}{P_{Z}(Z_{i})}
\]
concentrates fast around its expected value 
\[
I(Z^{n};X^{n}=x^{n}):=\sum_{i=1}^{n}\E\left[\log\frac{P_{Z\mid X}(Z_{i}\mid x_{i})}{P_{Z}(Z_{i})}\,\middle\vert\,X_{i}=x_{i}\right].
\]
We achieve this using a concentration bound of Lipschitz functions
of Poisson RVs due to Bobkov and Ledoux \cite[Prop. 11]{bobkov1998modified}
stated in Lemma \ref{lem: Poisson concentration} (Appendix \ref{sec:Poisson-concentration-of}).
The result is as follows: 
\begin{lem}
\label{lem: concentration of information density wrt z}Assume that
$\supp(P_{X})\subseteq[s_{n}]$ for some $s_{n}\in\mathbb{N}_{+}$.
Let $x^{n}\in([s_{n}])^{\otimes n}$. Then, for any $\delta\in(0,\frac{r_{n}}{g_{n}}s_{n})$\textup{
\[
\P\left[f_{x^{n}}(Z^{n})<I(Z^{n};X^{n}=x^{n})-n\delta\mid X^{n}=x^{n}\right]\leq\exp\left[-n\frac{g_{n}\delta^{2}}{19r_{n}s_{n}\log^{2}s_{n}}\right].
\]
}
\end{lem}
\begin{IEEEproof}
To establish this, we begin by showing that if $x^{n}\in([s_{n}])^{\otimes n}$
then $f_{x^{n}}(z^{n})$ is Lipschitz with semi-norm $\beta=\log s_{n}$,
as follows. We denote by $e^{n}(i)=(0,0,\ldots,1,0..)$ the $i$th
standard basis vector in $\mathbb{R}^{n}$. Let $Z\mid X=x\sim\Poi(\frac{r_{n}}{g_{n}}x)$.
Then, it holds for any $x\in\mathbb{R}_{+}$ and $z\in\mathbb{N}$
that 
\begin{align}
\frac{P_{Z\mid X}(z+1\mid x)}{P_{Z\mid X}(z\mid x)} & =\frac{e^{-r_{n}x/g_{n}}\left(\frac{r_{n}x}{g_{n}}\right)^{z+1}}{(z+1)!}\cdot\frac{z!}{e^{-r_{n}x/g_{n}}\left(\frac{r_{n}x}{g_{n}}\right)^{z}}\\
 & =\frac{r_{n}}{g_{n}}\frac{x}{z+1}.
\end{align}
Let $P_{Z}$ be the marginal resulting from $P_{X}\otimes P_{Z\mid X}$.
Then, similarly, 
\begin{align}
\frac{P_{Z}(z)}{P_{Z}(z+1)} & =\frac{\sum_{\tilde{x}\in\supp(P_{X})}P_{X}(\tilde{x})P_{Z\mid X}(z\mid\tilde{x})}{\sum_{\tilde{x}\in\supp(P_{X})}P_{X}(\tilde{x})P_{Z\mid X}(z+1\mid\tilde{x})}\\
 & \leq\max_{\tilde{x}\in\supp(P_{X})}\frac{P_{Z\mid X}(z\mid\tilde{x})}{P_{Z\mid X}(z+1\mid\tilde{x})}\\
 & =\max_{\tilde{x}\in\supp(P_{X})}\frac{g_{n}}{r_{n}}\frac{z+1}{\tilde{x}}.
\end{align}
Hence, 
\[
\frac{P_{Z\mid X}(z+1\mid x)}{P_{Z\mid X}(z\mid x)}\frac{P_{Z}(z)}{P_{Z}(z+1)}\leq\max_{\tilde{x}\in\supp(P_{X})}\frac{x}{\tilde{x}}\leq s_{n}.
\]
Analogously, we can prove that 
\[
\frac{P_{Z\mid X}(z+1\mid x)}{P_{Z\mid X}(z\mid x)}\frac{P_{Z}(z)}{P_{Z}(z+1)}\geq\min_{\tilde{x}\in\supp(P_{X})}\frac{x}{\tilde{x}}\geq1.
\]
Thus, for any $x\in\supp(P_{X})\subseteq[s_{n}]$ and $z\in\mathbb{N}$
\[
\left|\log\frac{P_{Z\mid X}(z+1\mid x)}{P_{Z}(z+1)}-\log\frac{P_{Z\mid X}(z\mid x)}{P_{Z}(z)}\right|\leq\log s_{n}.
\]
The additive form of $f_{x^{n}}(z^{n})$ then implies that \emph{
\begin{equation}
\max_{z^{n}\in\mathbb{N}^{n}}\left|f_{x^{n}}(z^{n}+e^{n}(i))-f_{x^{n}}(z^{n})\right|\leq\log s_{n}.\label{eq: Lipschitz constant of information density}
\end{equation}
}

This Lipschitz property results in a left-tail concentration of $f_{x^{n}}(Z^{n})$,
by invoking a variant of the Bobkov--Ledoux concentration inequality\cite[Prop. 11]{bobkov1998modified}
(see Lemma \ref{lem: Poisson concentration}) on the function $-f_{x^{n}}(Z^{n})$
of the Poisson RVs $Z_{i}\mid X_{i}=x\sim\Poi(\frac{r_{n}}{g_{n}}x_{i})$.
Specifically, since (\ref{eq: Lipschitz constant of information density})
implies that $f_{x^{n}}(z^{n})$ is Lipschitz with semi-norm $\beta=\log s_{n}$,
Lemma \ref{lem: Poisson concentration} results
\[
\P\left[f_{x^{n}}(Z^{n})-I(Z^{n};X^{n}=x^{n})<-n\delta\mid X^{n}=x^{n}\right]\leq\exp\left[-n\cdot\frac{\delta^{2}}{16\beta^{2}\overline{\lambda}+3\beta\delta}\right],
\]
where $\overline{\lambda}\leq\max_{i\in[n]}\frac{r_{n}}{g_{n}}x_{i}\leq\frac{r_{n}}{g_{n}}s_{n}$.
The concentration result stated in the lemma then follows by utilizing
the assumption that $x_{i}\in[s_{n}]$ for all $i\in[n]$, and by
slightly loosening the bound, using the assumption $\delta\leq\frac{r_{n}}{g_{n}}s_{n}$.
\end{IEEEproof}
We continue to the second step in analyzing the probability in (\ref{eq: Poissonization of the information density CDF}),
which is the analysis of the randomness of $X^{n}$. To this end,
we denote
\begin{align}
J(x^{n}) & :=\sum_{i=1}^{n}\E\left[\log P_{Z\mid X}(Z_{i}\mid x_{i})\,\middle\vert\,X_{i}=x_{i}\right]\\
 & =\sum_{i=1}^{n}I(Z_{i};X_{i}=x_{i})-H(Z_{i}).
\end{align}
\begin{lem}
\label{lem: concentration of information density wrt x}Assume that
$\supp(P_{X})\subseteq[s_{n}]$, and $r_{n}s_{n}\geq12\pi e^{2}g_{n}$.
Then, 

\[
\P\left[J(X^{n})+H(Z^{n})<I(X^{n};Z^{n})-n\delta\right]\leq\exp\left[-n\cdot\frac{2\delta^{2}}{\log^{2}\frac{r_{n}s_{n}}{g_{n}}}\right].
\]
\end{lem}
\begin{IEEEproof}
We note that $\E[J(X^{n})]=-H(Z^{n}\mid X^{n})$, and show that $J(X^{n})$
concentrates to its expected value using Hoeffding's inequality (Fact
\ref{fact: Hoeffding inequality}, Appendix \ref{sec:Useful-Mathematical-Results}).
We begin by noting that it holds that 
\begin{equation}
\E\left[\log P_{Z\mid X}(Z_{i}\mid x_{i})\mid X_{i}=x_{i}\right]\leq0.\label{eq: upper bound on the expected log likelihood}
\end{equation}
Also, 
\begin{align}
-\log P_{Z\mid X}(z\mid x) & =\log\frac{z!}{e^{-r_{n}x/g_{n}}\left(\frac{r_{n}x}{g_{n}}\right)^{z}}\\
 & \leq\frac{r_{n}x}{g_{n}}+\log z!+z\cdot\log\left(\frac{g_{n}}{r_{n}x}\right).
\end{align}
For $z\geq1$, using Stirling's bound (Fact \ref{fact: Stirling bound}
in Appendix \ref{sec:Useful-Mathematical-Results})
\[
\log z!\leq z\log z-z+\frac{1}{2}\log(6\pi z),
\]
and for $z=0$ it holds that $\log z!=0$. Hence, for $Z\mid X=x\sim\Poi(\frac{r_{n}}{g_{n}}x)$
\begin{align}
 & \E\left[\log Z!\mid X=x\right]\nonumber \\
 & \leq\E\left[\left(Z\log Z-Z+\frac{1}{2}\log(6\pi Z)\right)\cdot\I\{Z>0\}\,\middle\vert\,X=x\right]\\
 & \trre[\leq,a]\E\left[Z\log Z-Z+\frac{1}{2}\log(6\pi(Z+1))\,\middle\vert\,X=x\right]\\
 & \trre[\leq,b]\frac{r_{n}x}{g_{n}}\log\left(1+\frac{r_{n}x}{g_{n}}\right)-\frac{r_{n}x}{g_{n}}+\E\left[\frac{1}{2}\log(6\pi(Z+1))\,\middle\vert\,X=x\right]\\
 & \trre[\leq,c]\frac{r_{n}x}{g_{n}}\log\left(1+\frac{r_{n}x}{g_{n}}\right)-\frac{r_{n}x}{g_{n}}+\frac{1}{2}\log\left(6\pi\left(\frac{r_{n}x}{g_{n}}+1\right)\right),\label{eq: upper bound on expected log Z factorial}
\end{align}
where $(a)$ follows by analytically completing $Z\log Z=0$ for $Z=0$,
and upper bounding $\log Z\leq\log(Z+1)$, $(b)$ follows since if
$V\sim\Poi(\lambda)$ then, $\E[V\log V]\leq\lambda\log(1+\lambda)$
(see Lemma \ref{lem: expectation of VlogV for Poisson} in Appendix
\ref{sec:The-Poisson-Distribution} for a proof), $(c)$ follows from
Jensen's inequality for the concave logarithm function. So, 
\begin{align}
 & \E\left[-\log P_{Z\mid X}(Z\mid x_{i})\mid X=x_{i}\right]\\
 & =\frac{r_{n}x_{i}}{g_{n}}+\E\left[\log Z!\mid X=x_{i}\right]+\E\left[Z\mid x=x_{i}\right]\cdot\log\left(\frac{g_{n}}{r_{n}x_{i}}\right)\\
 & =\frac{r_{n}x_{i}}{g_{n}}+\E\left[\log Z!\mid X=x_{i}\right]+\frac{r_{n}x_{i}}{g_{n}}\log\left(\frac{g_{n}}{r_{n}x_{i}}\right)\\
 & \trre[\leq,a]\frac{r_{n}x_{i}}{g_{n}}\left[1+\log\left(\frac{g_{n}}{r_{n}x_{i}}\right)\right]+\frac{r_{n}x_{i}}{g_{n}}\log\left(1+\frac{r_{n}x_{i}}{g_{n}}\right)-\frac{r_{n}x_{i}}{g_{n}}+\frac{1}{2}\log\left(6\pi\left(\frac{r_{n}x_{i}}{g_{n}}+1\right)\right)\\
 & =\frac{r_{n}x_{i}}{g_{n}}\log\left(1+\frac{g_{n}}{r_{n}x_{i}}\right)+\frac{1}{2}\log\left(6\pi\left(\frac{r_{n}x_{i}}{g_{n}}+1\right)\right)\\
 & \trre[\leq,b]1+\frac{1}{2}\log\left(6\pi\left(\frac{r_{n}x_{i}}{g_{n}}+1\right)\right)\\
 & \trre[\leq,c]\frac{1}{2}\log\left(6\pi e^{2}\left(\frac{r_{n}s_{n}}{g_{n}}+1\right)\right)\\
 & \trre[\leq,d]\frac{1}{2}\log\left(12\pi e^{2}\frac{r_{n}s_{n}}{g_{n}}\right),\\
 & \trre[\leq,e]\log\frac{r_{n}s_{n}}{g_{n}},\label{eq: upper bound on the minus expected log likelihood}
\end{align}
where $(a)$ follows from (\ref{eq: upper bound on expected log Z factorial}),
$(b)$ follows from $\log(1+t)\leq t$, $(c)$ follows since under
the assumption of the lemma $x_{i}\leq s_{n}$, and both $(d)$ and
$(e)$ follow by the assumption $\frac{r_{n}s_{n}}{g_{n}}\geq12\pi e^{2}\geq1$. 

We deduce from (\ref{eq: upper bound on the expected log likelihood})
and (\ref{eq: upper bound on the minus expected log likelihood})
that $J(X^{n})$ is a sum of $n$ independent RVs $\E_{Z_{i}\mid X_{i}}[\log P_{Z\mid X}(Z_{i}\mid X_{i})]$,
each of which is bounded, with probability $1$, in $[-\log\frac{r_{n}s_{n}}{g_{n}},0].$
Consequently, Hoeffding's inequality (Fact \ref{fact: Hoeffding inequality}
in Appendix \ref{sec:Useful-Mathematical-Results}) implies that 
\[
\P\left[J(X^{n})+H(Z^{n}\mid X^{n})<-n\delta\right]\leq\exp\left[-n\cdot\frac{2\delta^{2}}{\log^{2}\frac{r_{n}s_{n}}{g_{n}}}\right],
\]
which then implies the claim of the lemma, by adding $I(X^{n};Z^{n})$
to both sides in the inequality defining the event of interest.
\end{IEEEproof}
Setting $\delta_{n}\in(0,\frac{r_{n}}{g_{n}}s_{n})$, and then $\log\gamma=nI(X;Z)-2n\delta_{n}-\frac{1}{2}\log(6\pi nr_{n})$.
Let us define the event 
\[
{\cal E}_{n}(x^{n}):=\left\{ f_{x^{n}}(Z^{n})<I(Z^{n};X^{n}=x^{n})-n\delta_{n}\right\} .
\]
Then,
\begin{align}
 & \P\left[\sum_{i=1}^{n}\log\frac{P_{Z\mid X}(Z_{i}\mid X_{i})}{P_{Z}(Z_{i})}\leq\log\gamma+\frac{1}{2}\log(6\pi nr_{n})\right]\nonumber \\
 & =\P\left[\sum_{i=1}^{n}\log\frac{P_{Z\mid X}(Z_{i}\mid X_{i})}{P_{Z}(Z_{i})}\leq nI(X^{n};Z^{n})-2n\delta_{n}\right]\\
 & =\sum_{x^{n}\in[s_{n}]^{n}}\P[X^{n}=x^{n}]\cdot\P\left[f_{x^{n}}(Z^{n})\leq nI(X^{n};Z^{n})-2n\delta_{n}\,\middle\vert\,X^{n}=x^{n}\right]\\
 & \leq\sum_{x^{n}\in[s_{n}]^{n}}\P[X^{n}=x^{n}]\cdot\P\left[\left\{ f_{x^{n}}(Z^{n})\leq nI(X^{n};Z^{n})-2n\delta_{n}\right\} \cap\left\{ {\cal E}_{n}^{c}(x^{n})\right\} \,\middle\vert\,X^{n}=x^{n}\right]\nonumber \\
 & \hphantom{===}+\sum_{x^{n}\in[s_{n}]^{n}}\P[X^{n}=x^{n}]\cdot\P\left[{\cal E}_{n}(x^{n})\,\middle\vert\,X^{n}=x^{n}\right]\\
 & \trre[\leq,a]\sum_{x^{n}\in[s_{n}]^{n}}\P[X^{n}=x^{n}]\cdot\I\left[\left\{ I(Z^{n};X^{n}=x^{n})-n\delta_{n}\leq nI(X^{n};Z^{n})-2n\delta_{n}\right\} \right]\nonumber \\
 & \hphantom{===}+\sum_{x^{n}\in[s_{n}]^{n}}\P[X^{n}=x^{n}]\cdot\exp\left[-n\cdot\frac{2\delta_{n}^{2}}{\log^{2}\frac{r_{n}s_{n}}{g_{n}}}\right]\\
 & =\P\left[J(X^{n})+H(Z^{n})<I(X^{n};Z^{n})-n\delta\right]+\exp\left[-n\cdot\frac{2\delta_{n}^{2}}{\log^{2}\frac{r_{n}s_{n}}{g_{n}}}\right]\\
 & \trre[\leq,b]\exp\left[-n\cdot\frac{2\delta_{n}^{2}}{\log^{2}\frac{r_{n}s_{n}}{g_{n}}}\right]+\exp\left[-n\cdot\frac{g_{n}\delta_{n}^{2}}{19r_{n}s_{n}\log^{2}s_{n}}\right]\\
 & \leq2\exp\left[-n\delta_{n}^{2}\cdot\left(\frac{2}{\log^{2}\frac{r_{n}s_{n}}{g_{n}}}\wedge\frac{g_{n}}{19r_{n}s_{n}\log^{2}s_{n}}\right)\right],
\end{align}
where $(a)$ follows from the concentration bound in Lemma \ref{lem: concentration of information density wrt z},
and $(b)$ follows from Lemma \ref{lem: concentration of information density wrt x}.
We substitute this back into (\ref{eq: Poissonization of the information density CDF}),
and then in the extended Feinstein's bound (\ref{eq: extended Feinstein's bound})
to obtain 
\[
\epsilon_{n}P_{X}^{\otimes n}(F_{n})\leq4e\sqrt{nr_{n}}\exp\left[-n\delta_{n}^{2}\cdot\left(\frac{2}{\log^{2}\frac{r_{n}s_{n}}{g_{n}}}\wedge\frac{g_{n}}{19r_{n}s_{n}\log^{2}s_{n}}\right)\right]+\frac{M}{e^{nI(X^{n};Z^{n})-2n\delta_{n}-\frac{1}{2}\log(6\pi nr_{n})}}.
\]
The claim of the proposition is then proved by choosing $M=\exp\left[nI(X;Z)-3n\delta_{n}-\frac{1}{2}\log(6\pi nr_{n})\right]$,
and performing minor algebraic simplifications. 
\end{IEEEproof}
We now turn to prove Prop. \ref{prop: Truncated gamma MI}. 
\begin{IEEEproof}[Proof of Prop. \ref{prop: Truncated gamma MI}]
We analyze the reduction in mutual information over the Poisson channel,
resulting from modifying the ideal gamma distribution of $\tilde{X}$
to the truncated $\overline{X}:=\tilde{X}_{\mid{\cal S}_{n}}$, and
then to upward rounded $X:=\lceil\overline{X}\rceil$. We begin by
analyzing the reduction in mutual information due to the truncation
operation, using the I-MMPE relation (Theorem \ref{thm:I-MMPE}).
We begin with the following general result. 
\begin{lem}
\label{lem:truncation in the MI general}Let $U$ be a non-negative
RV, which satisfies $\E[U^{2}\log^{2}U]<\infty$, and let $\overline{U}\equiv U_{\mid{\cal S}}$
be distributed as $U$ truncated to an interval ${\cal S}:=[s_{\min},s_{\max}]\subset\mathbb{R}_{+}$,
as in Definition \ref{def: truncation} where $s_{\min}<1<s_{\max}$.
Let $a>0$ be given, assume that $V_{a}\mid U=u\sim\Poi(au)$, and
let $\overline{V}_{a}\mid\overline{U}=u\sim\Poi(au)$. Then, for any
$\gamma>0$
\begin{align}
I(U;V_{\gamma}) & \leq I(\overline{U};\overline{V}_{\gamma})+\gamma\cdot s_{\text{\emph{max}}}\cdot\P\left[U\in[0,s_{\text{\emph{min}}})\right]\nonumber \\
 & \hphantom{==}+\gamma\E\left[U\log U\cdot\I\left\{ U\in(s_{\text{\emph{max}}},\infty)\right\} \right]\nonumber \\
 & \hphantom{==}+\gamma\E\left[U\log\frac{1}{s_{\text{\emph{min}}}}\cdot\I\left\{ U\in(s_{\text{\emph{max}}},\infty)\right\} \right]
\end{align}
\end{lem}
\begin{IEEEproof}
Let $a>0$ be given. Let $\widehat{aU}(v)=\E[aU\mid V_{a}=v]$ be
the MMPE estimator of $aU$ based on the measurement $V_{a}$. Similarly,
let $\widehat{a\overline{U}}(v)=\E[a\overline{U}\mid\overline{V}_{a}=v]$
be the MMPE estimator of $a\overline{U}$ based on the measurement
$\overline{V}_{a}$. Recall that $\ell(u,v)=v-u+u\log\frac{u}{v}$
is Poisson error function. Then, 
\begin{align}
 & \mmpe(aU)\nonumber \\
 & =\E\left[\ell\left(aU,\widehat{aU}(V_{a})\right)\right]\\
 & \trre[\leq,a]\E\left[\ell\left(aU,\widehat{a\overline{U}}(V_{a})\right)\right]\\
 & \trre[=,b]a\E\left[\ell\left(U,\widehat{\overline{U}}(V_{a})\right)\right]\\
 & =a\E\left[\ell\left(U,\widehat{\overline{U}}(V_{a})\right)\,\middle\vert U\in{\cal S}\right]\cdot\P[U\in{\cal S}]+a\E\left[\ell\left(U,\widehat{\overline{U}}(V_{a})\right)\cdot\I\{U\not\in{\cal S}\}\right]\\
 & \trre[=,c]a\E\left[\ell\left(\overline{U},\widehat{\overline{U}}(\overline{V}_{a})\right)\right]\cdot\P[U\in{\cal S}]+a\E\left[\ell\left(U,\widehat{\overline{U}}(V_{a})\right)\cdot\I\{U\not\in{\cal S}\}\right]\\
 & \trre[\leq,d]\mmpe(a\overline{U})+a\E\left[\ell\left(U,\widehat{\overline{U}}(V_{a})\right)\cdot\I\{U\not\in{\cal S}\}\right]\\
 & =\mmpe(a\overline{U})+a\E\left[\ell\left(U,\widehat{\overline{U}}(V_{a})\right)\cdot\I\left\{ U\in[0,s_{\text{min}})\right\} \right]+a\E\left[\ell\left(U,\widehat{\overline{U}}(V_{a})\right)\cdot\I\left\{ U\in(s_{\max},\infty)\right\} \right]\\
 & \trre[\leq,e]\mmpe(a\overline{U})+a\E\left[\ell\left(U,s_{\text{max}}\right)\cdot\I\left\{ U\in[0,s_{\text{min}})\right\} \right]+a\E\left[\ell\left(U,s_{\text{min}}\right)\cdot\I\left\{ U\in(s_{\max},\infty)\right\} \right]\\
 & \trre[\leq,f]\mmpe(a\overline{U})+as_{\text{max}}\cdot\P\left[U\in[0,s_{\text{min}})\right]+a\cdot\E\left[U\log\frac{U}{s_{\text{min}}}\cdot\I\left\{ U\in(s_{\max},\infty)\right\} \right],\label{eq: bound on the MMPE in lemma}
\end{align}
where $(a)$ follows from the sub-optimality of $\widehat{a\overline{U}}$
for estimating $aU$, $(b)$ follows from the homogeneity property
of the loss function $\ell(au,av)=a\ell(u,v)$ for $a\geq0$, $(c)$
follows since conditioned on $U\in{\cal S}$, the distribution of
$U$ equals that of $\overline{U}$ and the distribution of $V_{a}$
equals that of $\overline{V}_{a}$, $(d)$ follows again from the
homogeneity property and $\P[U\in{\cal S}]\leq1$ and the term multiplying
$\P[U\in{\cal S}]$ is non-negative (the expected value of the Poisson
loss function), $(e)$ follows since $\widehat{\overline{U}}(V_{a})\in{\cal S}=[s_{\text{min}},s_{\text{max}}]$
and since the loss function $v\to\ell(u,v)$ is monotonic increasing
(resp. decreasing) for $v\geq u$ (resp. $v\leq u$), $(f)$ follows
since for $u\leq s_{\text{min}}<1$ it holds that 
\[
\ell\left(u,s_{\text{max}}\right)=s_{\text{max}}-u+u\log\frac{u}{s_{\text{max}}}\leq s_{\text{max}}+u\log\frac{u}{s_{\text{max}}}\leq s_{\text{max}},
\]
and for $u\geq s_{\text{max}}>1>s_{\text{min}}$ it holds that 
\[
\ell\left(u,s_{\text{min}}\right)=s_{\text{min}}-u+u\log\frac{u}{s_{\text{min}}}\leq s_{\text{min}}-u+u\log\frac{u}{s_{\text{min}}}\leq u\log\frac{u}{s_{\text{min}}}.
\]

Using twice the I-MMPE relation (Theorem \ref{thm:I-MMPE}), and the
bound (\ref{eq: bound on the MMPE in lemma}) directly leads to the
stated claim of the lemma, as 
\begin{align}
I(U;V_{\gamma}) & =\int_{0}^{\gamma}\mmpe(aU)\frac{\d a}{a}\\
 & \leq\int_{0}^{\gamma}\mmpe(a\overline{U})\frac{\d a}{a}+\nonumber \\
 & \hphantom{===}\int_{0}^{\gamma}\left\{ as_{\text{max}}\cdot\P\left[U\in[0,s_{\text{min}})\right]+a\cdot\E\left[U\log\frac{U}{s_{\text{min}}}\cdot\I\left\{ U\in(s_{\max},\infty)\right\} \right]\right\} \frac{\d a}{a}\\
 & \leq I(\overline{U};\overline{V}_{\gamma})+\gamma\cdot s_{\text{max}}\cdot\P\left[U\in[0,s_{\text{min}})\right]+\gamma\E\left[U\log\frac{U}{s_{\text{min}}}\cdot\I\left\{ U\in(s_{\max},\infty)\right\} \right].
\end{align}
\end{IEEEproof}
Lemma \ref{lem:truncation in the MI general} shows that the difference
in mutual information between the input $U$ and its truncated version
consists of three terms. Our next goal is to specifically evaluate
these terms for the distribution and support of interest, and show
that they are negligible $o_{n}(1)$. 
\begin{lem}
\label{lem: MI loss due to trunaction terms}Let $\tilde{X}\sim\Gam(\frac{1}{2},2g_{n})$,
and let $s_{\text{\emph{min}}}=\frac{1}{g_{n}^{1+3\rho}}$ and $s_{\text{\emph{max}}}=g_{n}^{1+\rho}$
for some $\rho\in(0,1)$. Then,
\begin{equation}
s_{\text{\emph{max}}}\cdot\P\left[\tilde{X}\in[0,s_{\text{\emph{min}}})\right]\leq\frac{1}{g_{n}^{\rho/2}}.\label{eq: MI loss trunaction first term}
\end{equation}
Also, there exists $n_{0}(\rho)$ such that for all $n\geq n_{0}(\rho)$
\begin{equation}
\E\left[\tilde{X}\log\tilde{X}\cdot\I\left\{ \tilde{X}\in(s_{\text{\emph{max}}},\infty)\right\} \right]\leq\exp\left[-\frac{g_{n}^{\rho}}{4}\right],\label{eq: MI loss trunaction second term}
\end{equation}
and 
\begin{equation}
\E\left[\tilde{X}\log\frac{1}{s_{\text{\emph{min}}}}\cdot\I\left\{ \tilde{X}\in(s_{\text{\emph{max}}},\infty)\right\} \right]\leq\exp\left[-\frac{g_{n}^{\rho}}{4}\right].\label{eq: MI loss trunaction third term}
\end{equation}
\end{lem}
\begin{IEEEproof}
We begin with the first term in (\ref{eq: MI loss trunaction first term}).
From the properties of the gamma probability distribution function
(PDF) of consideration (Lemma \ref{lem: Tails for Gamma 1/2} in Appendix
\ref{sec:Properties-of-the-Gamma-distribution}), it holds that 
\begin{align}
s_{\text{max}}\cdot\P\left[\tilde{X}\in[0,s_{\text{min}})\right] & =g_{n}^{1+\rho}\cdot\P\left[\tilde{X}\leq\frac{1}{g_{n}^{1+3\rho}}\right]\\
 & \leq\frac{g_{n}^{1+\rho}}{g_{n}^{(2+3\rho)/2}}\leq\frac{1}{g_{n}^{\rho/2}}.
\end{align}
We now move on to the second term in (\ref{eq: MI loss trunaction second term}).
For $t\in[g_{n}^{1+\rho},\infty)$, using the expression for the gamma
PDF (see Appendix \ref{sec:Properties-of-the-Gamma-distribution})
\begin{align}
\E\left[\tilde{X}\log\tilde{X}\cdot\I\left\{ \tilde{X}\geq g_{n}^{1+\rho}\right\} \right] & =\int_{g_{n}^{1+\rho}}^{\infty}\frac{1}{\sqrt{2\pi\tilde{x}g_{n}}}e^{-\tilde{x}/(2g_{n})}\cdot\tilde{x}\log\tilde{x}\cdot\d\tilde{x}\\
 & \trre[\leq,a]\int_{g_{n}^{1+\rho}}^{\infty}\frac{\tilde{x}}{\sqrt{2\pi g_{n}}}e^{-\tilde{x}/(2g_{n})}\cdot\d\tilde{x}\\
 & =\sqrt{\frac{2g_{n}}{\pi}}\int_{g_{n}^{1+\rho}}^{\infty}\frac{\tilde{x}}{2g_{n}}e^{-\tilde{x}/(2g_{n})}\cdot\d\tilde{x}\\
 & \trre[=,b]\frac{(2g_{n})^{3/2}}{\sqrt{\pi}}\int_{\frac{1}{2}g_{n}^{\rho}}^{\infty}se^{-s}\cdot\d s\\
 & \trre[=,c]\frac{(2g_{n})^{3/2}}{\sqrt{\pi}}\left(\frac{1}{2}g_{n}^{\rho}+1\right)e^{-\frac{1}{2}g_{n}^{\rho}}\\
 & \trre[\leq,d]4\exp\left[-\frac{1}{2}g_{n}^{\rho}+\left(\frac{3}{2}+\rho\right)\log g_{n}\right],
\end{align}
where $(a)$ follows since $\log\tilde{x}\leq\sqrt{\tilde{x}}$ for
$\tilde{x}\in\mathbb{R}_{+}$, $(b)$ is using the change of variables
$s=\frac{\tilde{x}}{2g_{n}}$, $(c)$ by solving the integral $\int se^{-s}\cdot\d s=-(s+1)e^{-s}$,
and $(d)$ follows since $g_{n}\geq1$. We finally move to the third
term in (\ref{eq: MI loss trunaction second term}). It holds that
\begin{align}
\E\left[\tilde{X}\cdot\I\left\{ \tilde{X}\geq g_{n}^{1+\rho}\right\} \right] & \trre[\leq,a]\E\left[\tilde{X}^{3/2}\cdot\I\left\{ \tilde{X}\in(s_{\max},\infty)\right\} \right]\\
 & \trre[\leq,b]\exp\left[-\frac{g_{n}^{\rho}}{3}\right],
\end{align}
where $(a)$ holds since $s_{\text{max}}=g_{n}^{1+\rho}\geq1$, and
$(b)$ holds as for the second term. The third term is bounded as
(\ref{eq: MI loss trunaction second term}) since 
\[
\log\frac{1}{s_{\text{min}}}=(1+3\rho)\log g_{n}\leq4\log g_{n}\leq\exp\left[-\frac{g_{n}^{\rho}}{12}\right].
\]
\end{IEEEproof}
Up until now we have considered the effect on the mutual information
of the Poisson channel when truncating the asymptotically optimal
input $\tilde{X}\sim\Gam(\frac{1}{2},2g_{n})$ to $\overline{X}=\tilde{X}_{\mid{\cal S}}$.
We next consider the effect on the mutual information of upward rounding
a continuous input $\overline{X}$ to an integer input $X=\lceil\overline{X}\rceil$.
To this end, we will decompose the mutual information $I(X;Z)=H(Z)-H(Z\mid X)$
and analyze how each of these terms changes due to the rounding operation. 

Let $\Psi(\mu):\mathbb{R}_{+}\to\mathbb{R}_{+}$ be the maximum entropy
for non-negative integer distributions, under a mean constraint $\mu$.
With a slight abuse of the notation in (\ref{eq: maximum entropy single letter})
we denote
\[
\Psi(\mu):=\max_{P_{A}}\left\{ H(A)\colon\supp(P_{A})\subseteq\mathbb{N}_{+},\;\E[A]\leq\mu\right\} .
\]
\begin{lem}
\label{lem:Geometric max entropy under mean constraint} $\Psi(\mu)=(\mu+1)\cdot\hbin\left(\frac{1}{\mu+1}\right).$
The function $\mu\to\Psi(\mu)$ is monotonic non-decreasing and concave
in $\mu$. 
\end{lem}
\begin{IEEEproof}
It is well known that the maximum entropy distribution among distributions
over the non-negative integers with a mean constraint is geometric.
For completeness, a standard proof is as follows. Let $A\sim P_{A}$
where $p_{i}:=P_{A}(i)$ for $i\in\mathbb{N}$. Assume that $P_{A}$
satisfies the mean constraint. Let $Q_{A}^{(\lambda)}$ be a distribution
defined by 
\[
q_{i}(\lambda)=\frac{e^{\lambda(\mu-i)}}{\sum_{j}e^{\lambda(\mu-j)}}
\]
for $i\in\mathbb{N}$. Then, for any $\lambda\geq0$
\begin{align}
H(P_{A}) & \leq\sum_{i=0}^{\infty}-p_{i}\log p_{i}+\lambda\left(\mu-\sum_{i=0}^{\infty}ip_{i}\right)\\
 & =\sum_{i=0}^{\infty}p_{i}\log\frac{e^{\lambda(\mu-i)}}{p_{i}}\\
 & =-\sum_{i=0}^{\infty}p_{i}\log\frac{p_{i}}{q_{i}(\lambda)}+\log\left(\sum_{j}e^{\lambda(\mu-j)}\right)\\
 & =-\Dkl(P_{A}\mid\mid Q_{A}^{(\lambda)})+\log\left(\sum_{j}e^{\lambda(\mu-j)}\right)\\
 & \leq\log\left(\sum_{j}e^{\lambda(\mu-j)}\right),
\end{align}
where equality holds if both $\E_{P}[A]=\sum_{i=0}^{\infty}ip_{i}=\mu$
and $P_{A}\equiv Q_{A}^{(\lambda)}$ holds. Now, $Q_{A}^{(\lambda)}$
is readily identified as a geometric distribution over $\mathbb{N}=\{0,1,2,..\}$.
Using the standard parametrization of the geometric distribution,
if $A\sim\Geo(\theta)$ then $\E[A]=\frac{1-\theta}{\theta}$ and
$H(A)=\frac{\hbin(\theta)}{\theta}$. Thus, $H(P_{A})=(\mu+1)\cdot\hbin(\frac{1}{\mu+1})$,
and this is the maximum entropy. Monotonicity is trivial, and concavity
is assured by the concavity of the entropy, or directly from the closed-form
expression of $\Psi(\mu)$. 
\end{IEEEproof}
Lemma \ref{lem:Geometric max entropy under mean constraint} will
next be used to compare the output entropy of the Poisson channel
when the input is a continuous $\overline{X}$, to that entropy when
the input is an integer rounded version of $\overline{X}$. We will
then also compare the conditional entropy $H(Z\mid X)$ under the
different input distributions can be easily compared, and combining
these two results we obtain a relation between the mutual information
values. Concretely: 
\begin{lem}
\label{lem: mutual information of  rounded gamma}Assume that $\underline{c}g_{n}\leq r_{n}\leq eg_{n}$
for some $\underline{c}\in(0,e)$ and let $\rho\in(0,1)$ be given,
and assume that $g_{n}\to\infty$ as $n\to\infty$. Let $\tilde{X}\sim\Gam(\frac{1}{2},2g_{n})$,
$\overline{X}=\tilde{X}_{\mid{\cal S}_{n}}$ with ${\cal S}_{n}=[g_{n}^{-(1+3\rho)},g_{n}^{1+\rho}]$
and let $X=\lceil\overline{X}\rceil$ be its upward rounding to the
nearest integer. Let $Z\sim\Poi(\frac{r_{n}}{g_{n}}X)$ and let $\overline{Z}\sim\Poi(\frac{r_{n}}{g_{n}}\overline{X})$.
Then, 
\[
I(X;Z)\geq I(\overline{X};\overline{Z})-\Psi\left(\frac{r_{n}}{g_{n}}\right)-o_{n}(1).
\]
\end{lem}
\begin{IEEEproof}
First, we compare the conditional entropy values $H(\overline{Z}\mid\overline{X})$
and $H(Z\mid X)$. Let $f_{\overline{X}}$ denote the density of $\overline{X}$
w.r.t. the Lebesgue measure $\lambda$, and let $P_{X}$ denote the
PMF of $X.$ We have that 
\[
H(\overline{Z}\mid\overline{X})=\int_{0}^{\infty}f_{\overline{X}}(\overline{x})\cdot H(\overline{Z}\mid\overline{X}=\overline{x})\cdot\lambda(\d\overline{x})
\]
where $H(\overline{Z}\mid\overline{X}=\overline{x})$ is the entropy
of a Poisson RV, and, similarly, 
\[
H(Z\mid X)=\sum_{i=0}^{\infty}P_{X}(i)\cdot H(Z\mid X=i).
\]
We show that the contribution to this sum by ``small'' indices $i\in[\lfloor g_{n}^{1-\rho}\rfloor]$
is negligible. Indeed, by Lemma \ref{lem: Poisson entropy}, it holds
that for any $i\in[\lfloor g_{n}^{1-\rho}\rfloor]$
\begin{align}
H(Z\mid X=i) & \leq\frac{1}{2}\log\left[2\pi e\left(i\frac{r_{n}}{g_{n}}+\frac{1}{12}\right)\right]\\
 & \leq\frac{1}{2}\log\left[2\pi e\left(2\frac{r_{n}}{g_{n}^{\rho}}+\frac{1}{12}\right)\right]\\
 & \leq\frac{1}{2}\log(35r_{n})\label{eq: Conditional entropy when  X is very small}
\end{align}
assuming that $\rho$ is sufficiently small so that $2\frac{r_{n}}{g_{n}^{\rho}}\geq\frac{1}{12}$.
Now, Lemma \ref{lem: Tails for Gamma 1/2} (Appendix \ref{sec:Properties-of-the-Gamma-distribution})
implies that it holds that 
\begin{align}
\P[\tilde{X}\not\in{\cal S}_{n}] & =\P[\tilde{X}\leq g_{n}^{-(1+3\rho)}]+\P[\tilde{X}\geq g_{n}^{1+\rho}]\\
 & \leq\frac{1}{g_{n}^{1+3\rho/2}}+2e^{-g_{n}^{\rho}/2}\\
 & \leq\frac{1}{2}
\end{align}
for all $n\geq n_{0}(\rho)$. Hence, using again (\ref{eq: lower tail of gamma 1/2})
Lemma \ref{lem: Tails for Gamma 1/2}, it holds that 
\begin{align}
\P\left[X\leq\lfloor g_{n}^{1-\rho}\rfloor\right] & =\P\left[\overline{X}\leq\lfloor g_{n}^{1-\rho}\rfloor\right]\\
 & \leq\frac{\P[\tilde{X}\leq g_{n}^{1-\rho}]}{\P[\tilde{X}\in{\cal S}_{n}]}\\
 & \leq\frac{\nicefrac{1}{g_{n}^{\rho/2}}}{\P[\tilde{X}\in{\cal S}_{n}]}\\
 & \leq\frac{2}{g_{n}^{\rho/2}}.\label{eq: Probabilty that X is very small for conditional entropy}
\end{align}
Furthermore, assume that $i\geq g_{n}^{1-\rho}$. Then, for any $\overline{x}\in[i-1,i]$
it holds from the asymptotic expression for the Poisson entropy in
Lemma \ref{lem: Poisson entropy} (Appendix \ref{sec:The-Poisson-Distribution})
that 
\begin{align}
 & H(Z\mid X=i)-H(Z\mid X=\overline{x})\\
 & =\frac{1}{2}\log\left[2\pi ei\frac{r_{n}}{g_{n}}\right]+O\left(\frac{1}{i\frac{r_{n}}{g_{n}}}\right)-\frac{1}{2}\log\left[2\pi e\overline{x}\frac{r_{n}}{g_{n}}\right]+O\left(\frac{1}{\overline{x}\frac{r_{n}}{g_{n}}}\right)\\
 & =\frac{1}{2}\log\left[\frac{i}{\overline{x}}\right]+O\left(\frac{1}{\overline{x}\frac{r_{n}}{g_{n}}}\right)\\
 & \leq\frac{1}{2}\log\left[\frac{\overline{x}+1}{\overline{x}}\right]+O\left(\frac{1}{\overline{x}\frac{r_{n}}{g_{n}}}\right)\\
 & =O\left(\frac{1}{\underline{c}\overline{x}}\right)\\
 & =O\left(\frac{1}{\underline{c}g_{n}^{1-\rho}}\right).\label{eq: conditional entropy difference for rounding input}
\end{align}
Thus, 
\begin{align}
H(Z\mid X) & =\sum_{i=0}^{\lfloor g_{n}^{1-\rho}\rfloor}P_{X}(i)\cdot H(Z\mid X=i)+\sum_{i=\lceil g_{n}^{1-\rho}\rceil}^{\infty}P_{X}(i)\cdot H(Z\mid X=i)\\
 & \trre[\leq,a]\P[X\leq\lfloor g_{n}^{1-\rho}\rfloor]\cdot H(Z\mid X=\lfloor g_{n}^{1-\rho}\rfloor)+\sum_{i=\lceil g_{n}^{1-\rho}\rceil}^{\infty}P_{X}(i)\cdot H(Z\mid X=i)\\
 & \trre[\leq,b]\frac{2}{g_{n}^{\rho/2}}\cdot\frac{1}{2}\log(35r_{n})+\sum_{i=\lceil g_{n}^{1-\rho}\rceil}^{\infty}P_{X}(i)\cdot H(Z\mid X=i)\\
 & =\frac{2}{g_{n}^{\rho/2}}\cdot\frac{1}{2}\log(35r_{n})+\sum_{i=\lceil g_{n}^{1-\rho}\rceil}^{\infty}\left[\int_{i-1}^{i}f_{\overline{X}}(\overline{x})\lambda(\d\overline{x})\right]\cdot H(Z\mid X=i)\\
 & \trre[\leq,c]\frac{2}{g_{n}^{\rho/2}}\cdot\frac{1}{2}\log(35r_{n})+\int_{\lceil g_{n}^{1-\rho}\rceil}^{\infty}f_{\overline{X}}(\overline{x})\cdot H(\overline{Z}\mid\overline{X}=\overline{x})\cdot\lambda(\d\overline{x})+O\left(\frac{1}{\underline{c}g_{n}^{1-\rho}}\right)\\
 & \leq o_{n}(1)+\int_{0}^{\infty}f_{\overline{X}}(\overline{x})\cdot H(\overline{Z}\mid\overline{X}=\overline{x})\cdot\lambda(\d\overline{x})\\
 & =o_{n}(1)+H(\overline{Z}\mid\overline{X}),\label{eq: conditional entropy by rounding}
\end{align}
where $(a)$ follows from the monotonicity of the Poisson entropy
as a function of its parameter, $(b)$ follows from (\ref{eq: Probabilty that X is very small for conditional entropy})
and (\ref{eq: Conditional entropy when  X is very small}), $(c)$
follows from (\ref{eq: conditional entropy difference for rounding input}). 

Second, we compare the output entropy $H(\overline{Z})$ and $H(Z)$.
To this end, we decompose $X=\overline{X}+D$ where $D\in[0,1]$ (note,
however, that $X$ and $D$ are statistically dependent). Conditioned
on $X=x$, or equivalently, on $(\overline{X},D)=(\overline{x},d)$,
it holds that $Z\sim\Poi(\frac{r_{n}}{g_{n}}(\overline{x}+d))$. By
the infinite divisibility of the Poisson distribution, we may write
$Z\eqd\overline{Z}+\acute{Z}$, where $\overline{Z}\sim\Poi(\frac{r_{n}}{g_{n}}\overline{x})$
and $\acute{Z}\sim\Poi(\frac{r_{n}}{g_{n}}d)$ are statistically independent.
We thus let $Z=\overline{Z}+\acute{Z}$, and then note that that $Z\geq\overline{Z}$,
with probability $1$, and that both are integer valued discrete RVs.
We continue similarly to the bound in \cite[Prop. 8]{polyanskiy2016wasserstein}.
It holds that 
\begin{align}
H(\overline{Z})-H(Z) & \leq H(Z,\overline{Z})-H(Z)\\
 & =H(\overline{Z}\mid Z)\\
 & =H(Z-\overline{Z}\mid Z)\\
 & \trre[\leq,a]\E\left[\Psi\left(\E\left[Z-\overline{Z}\mid Z\right]\right)\right]\\
 & \trre[\leq,b]\Psi\left(\E\left[Z-\overline{Z}\right]\right)\\
 & =\Psi\left(\frac{r_{n}}{g_{n}}\E\left[X-\overline{X}\right]\right)\\
 & \trre[\leq,c]\Psi\left(\frac{r_{n}}{g_{n}}\right),\label{eq: output entropy by rounding}
\end{align}
where $(a)$ follows from the operational definition of $\Psi(\mu)$,
$(b)$ follows since $\mu\to\Psi(\mu)$ is concave (Lemma \ref{lem:Geometric max entropy under mean constraint})
along with Jensen's inequality, and $(c)$ follows since $\mu\to\Psi(\mu)$
is monotonic non-decreasing in $\mu$ (Lemma \ref{lem:Geometric max entropy under mean constraint})
and as $0\leq X-\overline{X}\leq1$. 

Concluding, utilizing both (\ref{eq: conditional entropy by rounding})
and (\ref{eq: output entropy by rounding}) we obtain the claimed
bound. 
\end{IEEEproof}
We may now conclude the proof of Prop. \ref{prop: Truncated gamma MI}.
Let $\rho\in(0,1)$ be given, and let ${\cal S}_{n}=[g_{n}^{-(1+3\rho)},g_{n}^{1+\rho}]$.
Let $\tilde{X}\sim\Gam(\frac{1}{2},2g_{n})$, let $\overline{X}=\tilde{X}_{\mid{\cal S}_{n}}$,
and let $X=\lceil\overline{X}\rceil$. Let $\tilde{Z}\mid\tilde{X}=\tilde{x}\sim\Poi(\frac{r_{n}}{g_{n}}\tilde{x})$,
$\overline{Z}\mid\overline{X}=\overline{x}\sim\Poi(\frac{r_{n}}{g_{n}}\overline{x})$,
and $Z\mid X=x\sim\Poi(\frac{r_{n}}{g_{n}}x)$. It then holds for
all $n\geq n_{0}$ (which depends on $\underline{c}$ $\rho$, $\{g_{n}\}$),
\begin{align}
I(X;Z) & \trre[\geq,a]I(\overline{X};\overline{Z})-\Psi\left(\frac{r_{n}}{g_{n}}\right)-o_{n}(1)\\
 & \trre[\geq,b]I(\tilde{X};\tilde{Z})-\frac{r_{n}}{g_{n}}\cdot\left(\frac{1}{g_{n}^{\rho/2}}+2e^{-\frac{1}{4}g_{n}^{\rho}}\right)-\Psi\left(\frac{r_{n}}{g_{n}}\right)-o_{n}(1)\\
 & =I(\tilde{X};\tilde{Z})-\Psi\left(\frac{r_{n}}{g_{n}}\right)-o_{n}(1)\\
 & \trre[\geq,c]\frac{1}{2}\log r_{n}-\Psi\left(\frac{r_{n}}{g_{n}}\right)-o_{n}(1),
\end{align}
where $(a)$ follows from Lemma \ref{lem: mutual information of  rounded gamma},
$(b)$ follows from from Lemmas \ref{lem:truncation in the MI general}
and\ref{lem: MI loss due to trunaction terms}, and $(c)$ follows
from the known lower bound \cite[Thm. 7]{lapidoth2008capacity} on
the average-power-constrained Poisson channel.
\end{IEEEproof}
We continue with the proof of Prop. \ref{prop: Constant weight input high probability}: 
\begin{IEEEproof}[Proof of Prop. \ref{prop: Constant weight input high probability}]
Essentially, our goal is to analyze the probability that $\frac{1}{n}\sum_{i=1}^{n}X_{i}$
is significantly larger than its non-truncated mean (that is, the
mean of $\Gam(\frac{1}{2},2g_{n})$ distribution). Recall that $\{X_{i}\}$
are drawn from a \emph{rounded and truncated} Gamma distribution.
We next mainly discuss the truncation operation, as this has larger
effect on the analysis of that probability than the rounding operation.
Let us describe a generative model for RVs whose distribution is the
truncated gamma. Let $\{\tilde{X}_{i,j}\}_{i\in[n],j\in\mathbb{N}_{+}}$
be a double-index array of IID RVs, where $\tilde{X}_{i,j}\sim\Gam(\frac{1}{2},2g_{n})$.
Let 
\[
J^{*}(i):=\min\left\{ j\in\mathbb{N}_{+}\colon\tilde{X}_{i,j}\in{\cal S}_{n}\right\} 
\]
where ${\cal S}_{n}=[g_{n}^{-(1+3\rho)},g_{n}^{1+\rho}]$, and let
$\overline{X}_{i}=\tilde{X}_{J^{*}(i)}$. Then, $\overline{X}_{i}$
is distributed according to the $\Gam(\frac{1}{2},2g_{n})$ distribution,
truncated to ${\cal S}_{n}$, as required. Now, using Lemma \ref{lem: Tails for Gamma 1/2}
(Appendix \ref{sec:Properties-of-the-Gamma-distribution}), it holds
for any fixed $(i,j)$ that
\[
\P\left[\tilde{X}_{i,j}\not\in{\cal S}_{n}\right]=\P\left[\tilde{X}_{i,j}\leq g_{n}^{-(1+3\rho)}\right]+\P\left[\tilde{X}_{i,j}\geq g_{n}^{1+\rho}\right]\leq\frac{1}{g_{n}^{1+3\rho/2}}+2e^{-g_{n}^{\rho}/2}\leq\frac{2}{g_{n}^{1+3\rho/2}},
\]
where the last inequality holds for all $n$ sufficiently large. Let
$L:=\sum_{i=1}^{n}\I\{J^{*}(i)>1\}$ be the number of indices for
which $\tilde{X}_{i,1}\not\in{\cal S}_{n}$, and so also $\overline{X}_{i}\neq\tilde{X}_{i,1}$.
Hence, $\E[L]\leq\frac{2n}{g_{n}^{1+3\rho/2}}$, that is, $J^{*}(i)=1$
for almost all $i\in[n]$. More sharply, the event ${\cal G}:=\{L\geq\frac{3n}{g_{n}^{1+3\rho/2}}\}$
has low probability, and indeed, the relative Chernoff inequality
(setting $\xi=\frac{1}{2}$ in Fact \ref{fact: Relative Chernoff}
in Appendix \ref{sec:Useful-Mathematical-Results}) implies that
\[
\P[{\cal G}]=\P\left[L\geq\frac{3n}{g_{n}^{1+3\rho/2}}\right]\leq\exp\left[-\frac{1}{5}\frac{n}{g_{n}^{1+3\rho/2}}\right].
\]
So, letting $t>0$, we may decompose the probability of interest as
\begin{align}
\P\left[\frac{1}{n}\sum_{i=1}^{n}\overline{X}_{i}-g_{n}\geq t\right] & \leq\P\left[\left\{ \frac{1}{n}\sum_{i=1}^{n}\overline{X}_{i}-g_{n}\geq t\right\} \bigcap{\cal G}^{c}\right]+\P[{\cal G}]\\
 & \trre[\leq,a]\sum_{\ell=0}^{\lceil\frac{3n}{g_{n}^{1+3\rho/2}}\rceil}\P\left[\left\{ \frac{1}{n}\sum_{i=1}^{n}\overline{X}_{i}-g_{n}\geq t\right\} \bigcap\left\{ L=\ell\right\} \right]+\exp\left[-\frac{1}{5}\frac{n}{g_{n}^{1+3\rho/2}}\right]\\
 & \leq\sum_{\ell=0}^{\lceil\frac{3n}{g_{n}^{1+3\rho/2}}\rceil}\P\left[\left\{ \frac{1}{n}\sum_{i=1}^{n}\overline{X}_{i}-g_{n}\geq t\right\} \bigcap\left\{ L=\ell\right\} \right]+o_{n}(1),\label{eq: excess cost two events}
\end{align}
where $(a)$ follows from the assumption $n=\Omega(g_{n}^{1+\zeta})$
and $\rho\leq\frac{2\zeta}{3}$. We focus on a single term in the
summation above. Given that $L=\ell$ there are $\ell$ indices for
which $\overline{X}_{i}\neq\tilde{X}_{i,1}$. There are ${n \choose \ell}$
possible ways to choose those indices, and further conditioning on
one specific choice, all the conditional probabilities are the same.
Hence, 
\begin{align}
 & \P\left[\left\{ \frac{1}{n}\sum_{i=1}^{n}\overline{X}_{i}-g_{n}\geq t\right\} \bigcap\left\{ L=\ell\right\} \right]\nonumber \\
 & \trre[\leq,a]{n \choose \ell}\cdot\P\left[\left\{ \frac{1}{n}\left(\sum_{i=1}^{\ell}\overline{X}_{i}+\sum_{i=\ell+1}^{n}\tilde{X}_{i,1}\right)-g_{n}\geq t\right\} \cap\bigcap_{i=1}^{\ell}\{\overline{X}_{i}\neq\tilde{X}_{i,1}\}\cap\bigcap_{i=\ell+1}^{n}\{\overline{X}_{i}=\tilde{X}_{i,1}\}\right]\\
 & \leq{n \choose \ell}\cdot\P\left[\frac{1}{n}\left(\sum_{i=1}^{\ell}\overline{X}_{i}+\sum_{i=\ell+1}^{n}\tilde{X}_{i,1}\right)-g_{n}\geq t\right]\\
 & \trre[\leq,b]{n \choose \ell}\cdot\P\left[\frac{1}{n}\left(\sum_{i=\ell+1}^{n}\tilde{X}_{i,1}\right)+\ell\frac{g_{n}^{1+\rho}}{n}-g_{n}\geq t\right]\\
 & ={n \choose \ell}\cdot\P\left[\frac{1}{n-\ell}\sum_{i=\ell+1}^{n}\left(\tilde{X}_{i,1}-g_{n}\right)+\frac{\ell(g_{n}^{1+\rho}-g_{n})}{n-\ell}\geq\frac{n}{n-\ell}t\right]\\
 & \trre[\leq,c]{n \choose \ell}\cdot\P\left[\frac{1}{n-\ell}\sum_{i=\ell+1}^{n}\left(\tilde{X}_{i,1}-g_{n}\right)\geq t-1\right]\\
 & \trre[\leq,d]{n \choose \ell}\cdot\left\{ \exp\left[-\frac{(n-\ell)(t-1)}{4g_{n}}\right]+\exp\left[-\frac{(n-\ell)(t-1)^{2}}{8g_{n}^{2}}\right]\right\} \\
 & \trre[\leq,e]\exp\left[\frac{15n}{g_{n}^{1+3\rho/2}}\log(g_{n})\right]\cdot\left\{ \exp\left[-\frac{(n-\ell)(t-1)}{4g_{n}}\right]+\exp\left[-\frac{(n-\ell)(t-1)^{2}}{8g_{n}^{2}}\right]\right\} ,\label{eq: excess probability for truncated gamma for a given l}
\end{align}
where $(a)$ follows from the union bound, $(b)$ follows since $\overline{X}_{i}\in{\cal S}_{n}=[g_{n}^{-(1+3\rho)},g_{n}^{1+\rho}]$
with probability $1$, $(c)$ follows since $n=\Omega(g_{n})$ and
so 
\[
\frac{\ell(g_{n}^{1+\rho}-g_{n})}{n-\ell}\leq\frac{\frac{3n}{g_{n}^{1+3\rho/2}}g_{n}^{1+\rho}}{n-\frac{3n}{g_{n}^{1+3\rho/2}}}=\frac{3}{g_{n}^{\rho/2}-\frac{3}{g_{n}^{1+\rho}}}\leq1
\]
for all $n$ large enough (which depends on $\{g_{n}\}$) as $g_{n}\to\infty$,
$(d)$ holds since $\frac{1}{n-\ell}\sum_{i=\ell+1}^{n}\tilde{X}_{i,1}\sim\Gam(\frac{n-\ell}{2},\frac{2g_{n}}{n-\ell})$
so that $\E[\frac{1}{n-\ell}\sum_{i=\ell+1}^{n}\tilde{X}_{i,1}]=g_{n}$,
and using the tail inequality of sub-gamma RVs in Lemma \ref{lem: Tails for Gamma general}
(Appendix \ref{sec:Properties-of-the-Gamma-distribution}), $(e)$
follows since $\ell<\frac{n}{2}$, and as ${n \choose \ell}$ is monotonic
non-increasing for $\ell<\frac{n}{2}$; Then, by Stirling's bound
(Fact \ref{fact: binomial approximtion} in Appendix \ref{sec:Useful-Mathematical-Results})
\[
{n \choose \ell}\leq\exp\left[n\cdot\hbin\left(\frac{3}{g_{n}^{1+3\rho/2}}\right)\right]\leq\exp\left[\frac{15n}{g_{n}^{1+3\rho/2}}\log(g_{n})\right],
\]
where the last inequality assumes that $\rho\in(0,1)$, and uses $\hbin(t)\leq-2t\log t$
for $t\in[0,\frac{1}{2}]$, which is valid since $\frac{3}{g_{n}^{1+3\rho/2}}\leq\frac{1}{2}$
for all $n\geq n_{0}(\rho)$ as $g_{n}\to\infty$. So, choosing $t=g_{n}^{3/4+\rho}+1:=\overline{t}$
in (\ref{eq: excess probability for truncated gamma for a given l})
assures that 
\begin{align}
 & \P\left[\left\{ \frac{1}{n}\sum_{i=1}^{n}\overline{X}_{i}-g_{n}\geq\overline{t}\right\} \bigcap\left\{ L=\ell\right\} \right]\nonumber \\
 & \leq\exp\left[\frac{15n}{g_{n}^{1+3\rho/2}}\log(g_{n})\right]\cdot\left\{ \exp\left[-\frac{n(1-\frac{\ell}{n})g_{n}^{3/4+\rho}}{4g_{n}}\right]+\exp\left[-\frac{n(1-\frac{\ell}{n})g_{n}^{3/2+2\rho}}{8g_{n}^{2}}\right]\right\} \\
 & \leq\exp\left[-c\frac{n}{g_{n}^{1/4-\rho}}\right],
\end{align}
where the last inequality holds since $(1-\frac{\ell}{n})\to1$ as
$n\to\infty$ assuming $\ell\leq\lceil\frac{3n}{g_{n}^{1+3\rho/2}}\rceil=o(n)$.
Hence, from (\ref{eq: excess cost two events})
\[
\P\left[\left\{ \frac{1}{n}\sum_{i=1}^{n}\overline{X}_{i}-g_{n}\geq\overline{t}\right\} \bigcap{\cal G}^{c}\right]\leq\lceil\frac{3n}{g_{n}^{1+3\rho/2}}\rceil\cdot\exp\left[-c\frac{n}{g_{n}^{1/4-\rho}}\right]=o_{n}(1),
\]
which, by substituting back to \ref{eq: excess cost two events} then
implies that 
\[
\P\left[\frac{1}{n}\sum_{i=1}^{n}\overline{X}_{i}-g_{n}\geq\overline{t}\right]=o_{n}(1).
\]
Consequently, and as $\rho\in(0,1/4)$ was assumed, it holds for all
$n$ sufficiently large that 
\[
\sum_{i=1}^{n}\overline{X}_{i}\leq n\left(g_{n}+g_{n}^{3/4+\rho}+1\right)=:ng_{n}(1+\overline{\varsigma}_{n})
\]
with probability larger than $1/2$, where $\overline{\varsigma}_{n}:=g_{n}^{-1/4+\rho}+g_{n}^{-1}=o_{n}(1)$.
Now, the upward integer rounding implies that $X_{i}=\lceil\overline{X}_{i}\rceil$
for all $i\in[n]$. Letting $\varsigma_{n}:=g_{n}^{-1/4+\rho}+2g_{n}^{-1}=o_{n}(1)$,
it also holds that 
\[
\sum_{i=1}^{n}X_{i}\leq\sum_{i=1}^{n}(\overline{X}_{i}+1)\leq ng_{n}(1+\varsigma_{n})
\]
with probability larger than $1/2$. Since $\sum_{i=1}^{n}X_{i}$
is integer, there must exists $k\in\mathbb{N}$ such that $0\leq k\leq ng_{n}(1+\varsigma_{n})$
such that 
\[
\P\left[\sum_{i=1}^{n}X_{i}=k\right]\geq\frac{1}{2ng_{n}(1+\varsigma_{n})}\geq\frac{1}{3ng_{n}},
\]
for all $n$ large enough.
\end{IEEEproof}
We are now ready to prove the achievability bound of Theorem \ref{thm: noiseless kernel bounds}.
\begin{IEEEproof}[Proof of the achievability bound of Theorem \ref{thm: noiseless kernel bounds} ]
Recall the assumption $n=\Omega(g_{n}^{1+\zeta})$. Choose $\rho\in(0,\frac{1}{4}\wedge\frac{\zeta}{4})$,
and $\chi\in(0,1)$ and set $\underline{g}_{n}=\frac{g_{n}}{1+\chi}$.
Then, using Prop. \ref{prop: Constant weight input high probability}
for $\underline{g}_{n}$instead of $g_{n}$ implies that there exists
$\tau_{n}\in[g_{n}]$ such that 
\begin{equation}
P_{X^{n}}\left[F_{n}(\tau_{n})\right]\geq\frac{1}{3n\underline{g}_{n}}\label{eq: bound on probability of constraint set for theorem proof}
\end{equation}
for all $n$ sufficiently large. We also note that the input distribution
of $X$ used by Prop. \ref{prop: Constant weight input high probability}
is supported on $[1,s_{n}]$ with $s_{n}=\lceil g_{n}^{1+\rho}\rceil$.
Let us choose $\delta_{n}=g_{n}^{-\zeta/8}=o_{n}(1)$. Then, under
the theorem assumptions $\delta_{n}\in(0,\frac{r_{n}}{g_{n}}s_{n})$
as and so the condition of Prop. \ref{prop: Feinstein's bound for Poisson}
is fulfilled. It then implies that there is a codebook of cardinality
$M$ which satisfies 
\begin{align}
\frac{1}{n}\log M & \geq I(X;Z)-3\delta_{n}-\frac{1}{2n}\log(6\pi nr_{n})\\
 & \trre[\geq,a]\frac{1}{2}\log r_{n}-o_{n}(1)-\Psi\left(\frac{r_{n}}{\underline{g}_{n}}\right)\\
 & \trre[\geq,b]\frac{1}{2}\log r_{n}-o_{n}(1)-\Psi\left(\frac{r_{n}}{\underline{g}_{n}}\right),
\end{align}
where the last inequality follows from Prop. \ref{prop: Truncated gamma MI}.
At the same time, the maximal error probability of the codebook satisfies
\begin{align}
\epsilon_{n} & \trre[\leq,a]33n\underline{g}_{n}\left[\sqrt{nr_{n}}\exp\left[-n\delta_{n}^{2}\cdot\left(\frac{2}{\log^{2}\left(r_{n}\underline{g}_{n}^{\rho}\right)}\wedge\frac{1}{19r_{n}\underline{g}_{n}^{\rho}(1+\rho)^{2}\log^{2}\underline{g}_{n}}\right)\right]+e^{-n\delta_{n}}\right]\\
 & \trre[\leq,b]n^{4}\left[\exp\left[-c\frac{n\delta_{n}^{2}}{g_{n}^{1+2\rho}}\cdot\right]+e^{-n\delta_{n}}\right]\\
 & \trre[\leq,c]n^{4}\left[\exp\left[-cg_{n}^{\zeta-2\rho}\delta_{n}^{2}\cdot\right]+e^{-n\delta_{n}}\right]\\
 & \trre[\leq,d]n^{4}\left[\exp\left[-cg_{n}^{\zeta/2}\delta_{n}^{2}\cdot\right]+e^{-n\delta_{n}}\right]\\
 & \trre[=,e]o_{n}(1),
\end{align}
where $(a)$ is obtained from (\ref{eq: bound on probability of constraint set for theorem proof})
and setting $s_{n}=\underline{g}_{n}^{1+\rho}$, $(b)$ follows by
simplifying with $n\geq\underline{g}_{n}\geq\underline{g}_{n}^{\rho},$and
$n\geq\sqrt{r_{n}}$ as well as $n\geq33$ and $r_{n}\leq eg_{n}\leq e(1+\chi)g_{n}$,
which all hold for sufficiently large $n$, and some numerical constant
$c>0$, $(c)$ holds since $n=\Omega(g_{n}^{1+\zeta})$ for some $\zeta>0$,
$(d)$ holds due to the choice $\rho\leq\frac{\zeta}{4}$, and $(e)$
holds by the choice $\delta_{n}=g_{n}^{-\zeta/8}=o_{n}(1)$. The result
then follows by taking $n\to\infty$, and then $\chi\to0$. 
\end{IEEEproof}

\section{Conclusion and Future Research \label{sec:Conclusion-and-Future}}

In this paper, we have considered the capacity of frequency-based
channel with multinomial sampling, provided upper and lower bounds
on its capacity, and applied it to the log-cardinality scaling of
optimal DNA-storage codebooks in the short-molecule regime. There
are multiple avenues for future research. First, while our bounds
are rather tight, there is still a gap between the upper and lower
bound, and specifically, it is interesting to settle the optimal choice
of the normalized number of samples $r_{n}$. Second, the achievable
bound of Theorem \ref{thm: noiseless kernel bounds} is only applicable
under the condition $n=\Omega(g_{n}^{1+\zeta})$. This condition is
limiting, and specifically, it limits the application of the achievable
bound to the DNA storage channel to $\beta>\frac{1}{2\log|{\cal A}|}$,
that is, very short molecules are excluded. Inspecting the proof,
this condition stems from the concentration inequality for the information
spectrum in the Poisson channel in Prop. \ref{prop: Feinstein's bound for Poisson},
which results an upper bound on the concentration probability, for
which one of the terms is $\exp[-n\delta_{n}^{2}\cdot\frac{g_{n}}{19r_{n}s_{n}\log^{2}s_{n}}]$,
under the assumption that the input $X$ is supported on $[s_{n}]$.
However, in order for the truncation of the optimal input of the Poisson
channel to $[s_{n}]$ to have a negligible effect on the mutual information,
Prop. \ref{prop: Truncated gamma MI} requires that $s_{n}=\lceil g_{n}^{1+\rho}\rceil$
for some $\rho>0$. In turn, roughly speaking, the above probability
only decays when $n=\omega(g_{n})$, and this is the source of the
condition in the theorem. Consequently, possible removal of this condition
requires finding tighter bounds on the concentration of the information
spectrum of the Poisson channel, or a completely different approach.
Third, it is of interest to analyze noisy sequencing channels. Inspecting
the proof of Theorem \ref{thm: noiseless kernel bounds}, it appears
that this would require analyzing the capacity of a channel with input
$X^{n}$ and output $Z^{n}\mid X^{n}\sim\Poi(\hat{X}^{n}W_{n})$ (where
$\hat{X}^{n}$ is the normalized version of $X^{n}$). This is a Poisson
channel with non-standard memory between the symbols, that is, inter-symbol
interference \cite{martinez2005capacity}, or a multiple-input multiple-output
(MIMO) Poisson channel. Informally speaking, even if $W_{n}$ is a
invertible matrix, there are two differences compared to the noiseless
case. First, the input $X^{n}$ is still restricted as $\sum\hat{X}_{i}\leq1$,
and the channel $\hat{X}^{n}W_{n}$ may reduce this sum at the input
of the Poisson channel. Second, the achievable lower bound on the
capacity of the Poisson channel is obtained by lower bounding the
output entropy $H(Z^{n})$ with the differential entropy of the input
$H(Z^{n})\geq h(X^{n})$ \cite[Prop. 11]{lapidoth2008capacity}. Here,
$h(\hat{X}^{n}W_{n})$ has a reduced differential entropy by $\log\det W_{n}$,
which will further reduce the capacity. A rigorous analysis appears
challenging, and thus is left for future work. Finally, as common,
a more accurate analysis of the decay of the error probability $\epsilon_{n}$
and establishing a strong converse are also of interest. 

\appendices{\numberwithin{equation}{section}}

\section{Useful Mathematical Results \label{sec:Useful-Mathematical-Results}}
\begin{fact}[Stirling's bound]
\label{fact: Stirling bound} For $n\in\mathbb{N}$
\[
\sqrt{2\pi n}\left(\frac{n}{e}\right)^{n}\leq n!\leq\sqrt{2\pi en}\left(\frac{n}{e}\right)^{n}.
\]
\end{fact}
\begin{fact}
\label{fact: binomial approximtion} For $k_{n}=o(n)$ as $n\to\infty$,
it holds that
\begin{equation}
{n \choose k}\sim\left(\frac{ne}{k}\right)^{k}\frac{1}{\sqrt{2\pi k}}\exp\left(-\frac{k^{2}(1+o(1))}{2n}\right)\label{eq: binomial approximation small k}
\end{equation}
where $a_{n}\sim b_{n}$ means that $\lim_{n\to\infty}\frac{a_{n}}{b_{n}}=1$.
Also, 
\begin{equation}
{n \choose k}\leq2^{n\hbin(k/n)}.\label{eq: Stirling bound on binomial with entropy}
\end{equation}
\end{fact}
\begin{fact}[Hoeffding's inequality \cite{hoeffding1994probability}]
\label{fact: Hoeffding inequality}If $\{X_{i}\}_{i\in[n]}$ are
independent RVs and $a_{i}\leq X_{i}\leq b_{i}$ with probability
$1$ then 
\begin{equation}
\P\left[\sum_{i=1}^{n}X_{i}-\E[X_{i}]\geq t\right]\leq\exp\left[-\frac{2t^{2}}{\sum_{i=1}^{n}(b_{i}-a_{i})^{2}}\right].\label{eq: hoeffding inequality one sided}
\end{equation}
\end{fact}
\begin{fact}[The relative (multiplicative) Chernoff bound]
\label{fact: Relative Chernoff}For $B_{i}\sim\Ber(p)$ IID for $i\in[n]$
\begin{equation}
\P\left[\frac{1}{n}\sum_{i=1}^{n}B_{i}-p\geq\xi p\right]\leq\exp\left[-\frac{\xi^{2}p}{2+\xi}\right]\label{eq: relative Chernoff}
\end{equation}
for any $\xi>0$.
\end{fact}

\section{Properties of The Poisson Distribution \label{sec:The-Poisson-Distribution}}
\begin{fact}[Poissonization of the multinomial distribution]
\label{fact: Poissonization of the multinomial distribution}Let
$\tilde{M}\sim\Poi(M)$, and let $\tilde{G}$ be a random vector such
that $\tilde{G}\sim\Mul(\tilde{M},(p_{1},p_{2},\ldots p_{J}\})$ conditioned
on $\tilde{M}$, where $\sum_{j\in[J]}p_{j}=1$ and $p_{j}>0$. Then,
$\{\tilde{G}(j)\}_{j\in[J]}$ are statistically independent and $\tilde{G}(j)\sim\Poi(Mp_{j})$
(unconditioned on $\tilde{M}$). 
\end{fact}
Fact \ref{fact: Poissonization of the multinomial distribution} can
be verified by spelling out the conditional PMF of $\tilde{G}$ conditioned
on $\tilde{M}$ \cite[Thm. 5.6]{mitzenmacher2017probability} in case
$\{p_{j}\}$ are all equal, and can be easily extended to the non-uniform
case (as in, e.g., \cite[Lecture 11, Thm. 3.2]{Seshadhri2020lecture}).
The following then follows from \cite[Corollary 5.9]{mitzenmacher2017probability}:
\begin{lem}
\label{lem: Poissonization of events}Let $G\sim\Mul(M,(p_{1},p_{2},\ldots p_{J}\})$,
and let $\tilde{G}$ be an independent Poisson vector of the same
dimension so that $\E[\tilde{G}(j)]=\E[G(j)]=Mp_{j}$. Then, for any
event ${\cal E}$
\[
\P\left[G\in{\cal E}\right]\leq\sqrt{eM}\cdot\P\left[\tilde{G}\in{\cal E}\right].
\]
\end{lem}
\begin{lem}[{Chernoff's bound for Poisson RVs \cite[Thm. 5.4]{mitzenmacher2017probability}}]
\label{lem: Poisson chernoff} Let $Z\sim\Poi(\lambda)$. Then,
for $\alpha\leq1$
\begin{equation}
\P\left[Z\leq\alpha\lambda\right]\leq e^{-\lambda}\left(\frac{e}{\alpha}\right)^{\alpha\lambda}=e^{-\lambda(1-\alpha\log(e/\alpha))}\leq e^{-\frac{\lambda}{2}(1-\alpha)^{2}}.\label{eq: Poisson Chernoff general lower tail}
\end{equation}
\end{lem}
\begin{lem}[Poisson entropy]
\label{lem: Poisson entropy}Let $Z_{\lambda}\sim\Poi(\lambda)$.
Then,
\[
H(Z_{\lambda})=\frac{1}{2}\log\left[2\pi e\lambda\right]+O\left(\frac{1}{\lambda}\right).
\]
Also, 
\[
H(Z_{\lambda})\leq\frac{1}{2}\log\left[2\pi e\left(\lambda+\frac{1}{12}\right)\right].
\]
Finally, $H(Z_{\lambda})$ is monotonic non-decreasing in $\lambda$. 
\end{lem}
\begin{IEEEproof}
For the first properties, see \cite[Lemma 10, Lemma 17b, Lemma 19]{lapidoth2008capacity}.
For the monotonicity property, note that by the infinite divisibility
of the Poisson distribution, if $\lambda_{2}>\lambda_{1}$ then $Z_{\lambda_{2}}\eqd Z_{\lambda_{1}}+\breve{Z}$
where $Z_{\lambda_{1}}$ and $\breve{Z}\sim\Poi(\lambda_{2}-\lambda_{1})$
are independent. As conditioning reduces entropy
\[
H(Z_{\lambda_{2}})=H(Z_{\lambda_{1}}+\breve{Z})\geq H(Z_{\lambda_{1}}+\breve{Z}\mid\breve{Z})=H(Z_{\lambda_{1}}\mid\breve{Z})=H(Z_{\lambda_{1}}).
\]
\end{IEEEproof}
\begin{lem}
\label{lem: expectation of VlogV for Poisson}Let $V\sim\Poi(\lambda)$.
Then, 
\[
\E\left[V\log V\right]\leq\lambda\log(1+\lambda).
\]
\end{lem}
\begin{IEEEproof}
We follow the idea in \cite{321933}. For any $v>0$ and $u>0$ it
holds that $\log\frac{v}{u}\leq\frac{v}{u}-1$ and so 
\[
v\log v=v\log\frac{v}{u}+v\log u\leq\frac{v^{2}}{u}+v\log\frac{u}{e}.
\]
Hence, 
\begin{align}
\E\left[V\log V\right] & \leq\E\left[\frac{V^{2}}{u}+V\log\frac{u}{e}\right]\\
 & =\frac{\lambda+\lambda^{2}}{u}+\lambda\log\frac{u}{e}\\
 & =\lambda\log(1+\lambda),
\end{align}
when choosing $u=1+\lambda$. 
\end{IEEEproof}

\section{Properties of the Gamma Distribution \label{sec:Properties-of-the-Gamma-distribution}}

Let $X\sim\Gam(k,\theta)$ where $k>0$ and $\theta>0$. Then, the
PDF is 
\[
f_{\Gam}\left(x\mid k,\theta\right)=\frac{1}{\Gamma(k)\theta^{k}}x^{k-1}e^{-x/\theta}
\]
and the CDF is 
\[
F_{\Gam}\left(x\mid k,\theta\right)=\frac{1}{\Gamma(k)}\gamma(k,\frac{x}{\theta})
\]
where 
\[
\gamma\left(k,x\right):=\int_{0}^{x}t^{k-1}e^{-t}\d t
\]
 is the \emph{incomplete gamma function}. For the special case of
$k=\frac{1}{2}$ it holds that $\Gamma(\frac{1}{2})=\sqrt{\pi}$,
and 
\[
\gamma\left(\frac{1}{2},x\right)=\sqrt{\pi}\erf(\sqrt{x})=2\int_{0}^{\sqrt{x}}e^{-t^{2}}\d t,
\]
where $\erf(x)=1-2\Q(\sqrt{2}x)$ and $\Q(x)=\frac{1}{\sqrt{2\pi}}\int_{z}^{\infty}e^{-t^{2}/2}\d t$
is the Q-function (the tail distribution function of the standard
normal distribution).  Also recall that for $X\sim\Gam(k,\theta)$
it holds that $\E[X]=k\theta$ and $\V[X]=k\theta^{2}$.
\begin{lem}
\label{lem: Tails for Gamma 1/2}Let $X\sim\Gam(\frac{1}{2},2g_{n})$.
Then, for $\eta\in(-\infty,1)$
\begin{equation}
\P\left[X\leq g_{n}^{\eta}\right]\leq\frac{1}{g_{n}^{(1-\eta)/2}}\label{eq: lower tail of gamma 1/2}
\end{equation}
and for any $\rho\in(0,\infty)$
\[
\P\left[X\geq g_{n}^{1+\rho}\right]\leq2e^{-g_{n}^{\rho}/2}.
\]
Thus, 
\[
\P\left[X\not\in[g_{n}^{\rho},g_{n}^{1+\rho}]\right]\leq[1+o_{n}(1)]\cdot\frac{1}{g_{n}^{(1-\rho)/2}}.
\]
\end{lem}
\begin{IEEEproof}
It holds that 
\begin{align}
\P\left[X\leq g_{n}^{\eta}\right] & =F_{\Gam}\left(g_{n}^{\eta}\mid\frac{1}{2},2g_{n}\right)\\
 & =\frac{2}{\sqrt{\pi}}\int_{0}^{\sqrt{\frac{g_{n}^{\eta}}{2g_{n}}}}e^{-t^{2}}\d t\\
 & \leq\frac{2}{\sqrt{\pi}}\cdot\sqrt{\frac{g_{n}^{\eta}}{2g_{n}}}\\
 & \leq\frac{1}{g_{n}^{(1-\eta)/2}}.
\end{align}
Next, 
\begin{align}
\P\left[X\geq g_{n}^{1+\rho}\right] & \leq1-F_{\Gam}\left(g_{n}^{1+\rho}\mid\frac{1}{2},2g_{n}\right)\\
 & =1-\erf\left(\sqrt{\frac{g_{n}^{\rho}}{2}}\right)\\
 & =2\Q(g_{n}^{\rho/2})\\
 & \leq2e^{-g_{n}^{\rho}/2},
\end{align}
using Chernoff's bound on the Q-function. 
\end{IEEEproof}
\begin{lem}
\label{lem: Tails for Gamma general}Let $X\sim\Gam(k,\theta)$ where
$k>0$ and $\theta>0$. Then, for $t>0$
\begin{align}
\P\left[X\geq\E[X]+t\right] & =\P\left[X\geq k\theta+t\right]\\
 & \leq e^{-\frac{t}{2\theta}}+e^{-\frac{t^{2}}{4k\theta^{2}}}.
\end{align}
\end{lem}
\begin{IEEEproof}
It holds that $\V[X]=k\theta^{2}$. Now, \cite[Sec. 2.4]{boucheron2013concentration}
states that $X-\E[X]$ is sub-gamma RV on the right tail, with parameters
$(v,c)=(k\theta^{2},\theta)$. Hence, for any $s\geq0$
\[
\P\left[X-\E[X]\geq\sqrt{2k\theta^{2}s}+\theta s\right]\leq e^{-s}.
\]
Taking $t=\sqrt{2k\theta^{2}s}+\theta s$ we have that $t\leq2(\sqrt{2k\theta^{2}s}\vee\theta s)$
(a sum is less than twice the maximum), and so $s\geq\frac{t}{2\theta}\wedge\frac{t^{2}}{4k\theta^{2}}$.
Hence, 
\[
e^{-s}\leq\exp\left[-\left(\frac{t}{2\theta}\wedge\frac{t^{2}}{4k\theta^{2}}\right)\right]\leq e^{-\frac{t}{2\theta}}+e^{-\frac{t^{2}}{4k\theta^{2}}}.
\]
 
\end{IEEEproof}

\section{Poisson Concentration of Lipschitz Functions \label{sec:Poisson-concentration-of}}

Assume that $V\sim\Poi(\lambda)$. Then, Bobkov and Ledoux have shown
the following logarithmic Sobolev inequality \cite[Corollary 4]{bobkov1998modified}:
It holds for any strictly positive function $f\colon\mathbb{N}\to\mathbb{R}_{+}$
that 
\begin{align}
\Ent\left[f(V)\right] & :=\E\left[f(V)\log(f(V)\right]-\E\left[f(V)\right]\E\left[\log(f(V)\right]\\
 & \leq\lambda\E\left[\frac{1}{f(V)}\cdot\left|Df(V)\right|^{2}\right],
\end{align}
where $Df(v):=f(v+1)-f(v)$ for $v\in\mathbb{N}$ is the discrete
derivative. Consequently, they have shown that for any function $g\colon\mathbb{N}\to\mathbb{R}$
with $\max_{v\in\mathbb{N}}|Dg(v)|\leq\tau$ it holds that \cite[Eq. (24)]{bobkov1998modified}
\[
\Ent\left[e^{g(V)}\right]\leq\lambda e^{2\tau}\cdot\E\left[|Dg(V)|^{2}\cdot e^{g(V)}\right].
\]
In turn, this implies the following concentration result:
\begin{lem}[{Poisson concentration of Lipschitz functions, a variant of \cite[Prop. 11]{bobkov1998modified}}]
\label{lem: Poisson concentration}Let $V_{i}\sim\Poi(\lambda_{i})$
for $i\in[n]$ be independent, and let $\overline{\lambda}\geq\max_{i\in[n]}\lambda_{i}$
be given. Also let $f\colon\mathbb{N}^{n}\to\mathbb{R}$ be such that
\[
\max_{v^{n}\in\mathbb{N}^{n}}\left|f(v^{n}+e^{n}(i))-f(v^{n})\right|\leq\beta
\]
\textup{\emph{where $e^{n}(i)$ is the $i$th standard basis vector
in $\mathbb{R}^{n}$.}} Then, for any $t>0$
\begin{equation}
\P\left[f(V^{n})-\E[f(V^{n})]>n\delta\right]\leq\exp\left[-n\cdot\frac{\delta^{2}}{16\beta^{2}\overline{\lambda}+3\beta\delta}\right]\label{eq: Poisson concentration}
\end{equation}
\end{lem}
\begin{IEEEproof}
The condition in the lemma trivially implies that choosing $\alpha^{2}=n\beta^{2}$
results
\[
\sum_{i=1}^{n}\left|f(v^{n}+e^{n}(i))-f(v^{n})\right|^{2}\leq\alpha^{2}
\]
in the notation of \cite[Prop. 11]{bobkov1998modified}. The proof
therein then relies on the tensorization property (subadditivity)
of the entropy functional, which is stated for IID $\{V_{i}\}_{i\in[n]}$,
but holds more generally when they are just independent \cite[Thm.  4.22]{boucheron2013concentration}.
Then, since 
\[
\Ent\left[e^{g(V_{i})}\right]\leq\overline{\lambda}e^{2\tau}\cdot\E\left[|Dg(V_{i})|^{2}\cdot e^{g(V_{i})}\right],
\]
Herbst argument and the entropy method can be used in the exact same
manner to show that 
\begin{equation}
\P\left[f(V^{n})-\E[f(V^{n})]>n\delta\right]\leq\exp\left[-\frac{n\delta}{4\beta}\log\left(1+\frac{\delta}{2\beta\overline{\lambda}}\right)\right].\label{eq: Poisson concentration preliminary version}
\end{equation}
(we set $\alpha^{2}=n\beta^{2}$, $c_{1}=\overline{\lambda}$ and
$c_{2}=2$ in the bound therein). Finally, we note that
\begin{align}
u\log\left(1+u\right) & =(1+u)\log(1+u)-u+u-\log(1+u)\\
 & \trre[\geq,*]\frac{u^{2}}{2(1+u/3)}+u-\log(1+u)\\
 & \trre[\geq,**]\frac{u^{2}}{2(1+u/3)},
\end{align}
where $(*)$ was stated in \cite[Exercise 2.8]{boucheron2013concentration},
and $(**)$ follows from $u\geq\log(1+u)$ for $u\geq0$. Using this
bound in (\ref{eq: Poisson concentration preliminary version}) with
$u=\frac{\delta}{2\beta\overline{\lambda}}$ establishes the claim
of the lemma. 
\end{IEEEproof}
\bibliographystyle{ieeetr}
\bibliography{DNA_short}

\end{document}